\documentclass[aps,a4paper,showpacs,preprintnumbers,amsmath,amssymb,
  floatfix]{revtex4}
\usepackage{dcolumn}
\usepackage{bm}
\usepackage{amsmath,amssymb}
\usepackage{booktabs,subfigure}
\usepackage{graphicx,psfrag}
\usepackage{epsfig}
\usepackage{color} 
\usepackage{rotating}
\usepackage{slashed}
\usepackage{geometry}
\usepackage{rotating}
\usepackage{subfigure}
\usepackage{caption}
\usepackage{pstricks}
\usepackage{color}
\allowdisplaybreaks[3]
\newcommand{\be}{\begin{equation}}
\newcommand{\ee}{\end{equation}}
\newcommand{\bea}{\begin{eqnarray}}
\newcommand{\eea}{\end{eqnarray}}
\newcommand{\cw}[1][{}]{\ensuremath{\cos^{#1} \theta_{W}}}
\newcommand{\sw}[1][{}]{\ensuremath{\sin^{#1} \theta_{W}}}

\numberwithin{equation}{section} 
\usepackage{epsfig}
\usepackage{dcolumn} 
\usepackage{bm} 

\def\gsim{\lower0.5ex\hbox{$\:\buildrel >\over\sim\:$}}
\def\lsim{\lower0.5ex\hbox{$\:\buildrel <\over\sim\:$}}

\begin{document}

\title{Invisible decays of the lightest Higgs boson in supersymmetric models}

\vskip 2cm

\author{B.Ananthanarayan,$^1$
~Jayita Lahiri,$^1$
~P. N. Pandita,$^2$
~Monalisa Patra$^1$}

\affiliation{$^1$ Centre for High Energy Physics, 
                  Indian Institute of Science, Bangalore 560 012, India \\
$^2$ Department of Physics, North Eastern Hill University, 
                  Shillong 793 002, India %
             }
\thispagestyle{myheadings}


\vskip 5cm

\begin{abstract}
\noindent
We consider supersymmetric models in which the lightest Higgs scalar can decay
invisibly 
consistent 
with the constraints on the $126$~GeV state discovered at the CERN LHC.
We  consider the invisible decay in the minimal supersymmetric 
standard model~(MSSM), as well
its extension containing an additional chiral singlet  superfield, the so-called
next-to-minimal or nonminimal supersymmetric standard model~(NMSSM).
We consider the case of MSSM with both universal as well as  
nonuniversal gaugino masses at the grand unified scale,
and find that only an $E_6$ grand unified 
model with unnaturally large representation can give rise to
sufficiently light neutralinos which can possibly lead to the 
invisible decay $h^0 \rightarrow \tilde \chi_1^0 \tilde \chi_1^0$. 
Following this, we consider the case of NMSSM in  detail, where also 
we find that it is not possible to have the invisible decay of the 
lightest Higgs scalar
with universal gaugino masses at the grand unified scale.
We delineate the regions of the NMSSM parameter space where it is
possible to have the lightest Higgs boson to have a mass of about
$126$ GeV, and then concentrate on the  region where this Higgs can decay 
into light neutralinos, with the soft gaugino masses
$M_1$ and $M_2$ as two independent parameters, unconstrained 
by grand unification.  We also consider, simultaneously,
the other important invisible Higgs decay channel
in the NMSSM,  namely the decay into the lightest CP odd scalars,
$h_1 \to a_1 a_1$, which is  studied in detail.  
With the invisible Higgs branching ratio 
being constrained by the present LHC results, we find that 
$\mu_{eff} < 170$~GeV and $M_1 < 80$~GeV is disfavored
in NMSSM for fixed values of the other input parameters. 
The dependence of our results
on the parameters of NMSSM is  discussed in detail.
 
\end{abstract}

\maketitle

\section{Introduction}
\label{sec:intro}

There is now a possible signal for a Higgs boson at a mass 
of around 126 GeV from the ATLAS~\cite{arXiv:1207.7214, ATLAS:2013mma} and
CMS~\cite{arXiv:1207.7235, Chatrchyan:2013lba} collaborations.  Attention is focused on
to check whether the decay widths of this particle are in accordance with 
the predictions of the Standard Model (SM) or its extensions,
especially the supersymmetric extensions of the SM.  
It may, however, turn out that the SM is
only a low-energy effective theory and that there are indeed particles 
of low masses that have evaded detection in the past due to their weak
coupling to the SM particles.  Candidates include such particles as 
the lightest neutralino in the minimal supersymmetric~(MSSM) extension
of the SM, and also the lightest 
CP-odd neutral Higgs boson of the next-to-minimal or 
nonminimal supersymmetric standard model~(NMSSM). 
The Higgs sector in MSSM is extended compared to the SM and
includes two Higgs doublets $H_1,{\rm and}~H_2$ leading to five 
physical Higgs states, which include two CP even Higgs bosons
$h$ and $H$ ($m_h < m_H$), a CP odd Higgs, $A$, and a pair
of charged Higgs bosons, $H^\pm$.  The recent discovery of the
Higgs like particle (with mass $m_h \approx 126$ GeV) at the
LHC, requires a significant degree of fine-tuning in the parameters 
in the context of MSSM. This fine-tuning can be evaded in case of the NMSSM, 
which is a extension of the  MSSM, supplemented by a chiral
singlet superfield~($S$). In the NMSSM the role of the $\mu$
parameter of the MSSM is played  by $\lambda <S>$, which is 
generated from a trilinear superpotential coupling 
$\lambda H_1H_2S$, when $S$ obtains a vacuum expectation value
$<S>$. This in turn leads to three CP-even Higgs bosons, $h_{1,2,3}$,
two CP-odd Higgs bosons, $a_{1,2},$ and a pair of charged 
Higgs bosons, $H^\pm$.
The  existence of the singlet chiral superfield not only 
has implications for the Higgs sector, but also
for the  neutralino sector, where the spectrum 
has an additional state when compared to the 
neutralino sector of the MSSM.
It has been found that certain regions of the parameter space
of MSSM  allow a Higgs boson ($h$) with a mass of $126$ GeV, albeit 
with fine tuning, satisfying the LHC results.


Since the identification of  the state with mass of $126$ GeV at the
LHC with the Higgs boson depends on the measurement
of its couplings to different particles, 
it is important to study all its  decay 
channels in the context of the SM and its supersymmetric extensions.
In the allowed parameter space there are regions where the Higgs
decay to the lightest neutralinos is kinematically allowed.
This in turn will lead to  invisible decay modes. 
Detailed studies have been carried out, where by assuming the discovered
particle to be the  SM Higgs boson, global fits have been performed
to place upper bounds on its invisible decay width. The
fits are performed for several cases, $(a)$ with the
assumption that the invisible Higgs width is the only new physics;
$(b)$ the couplings of Higgs to gluons and photons are considered as free
parameters, keeping the couplings to fermions and vector bosons to their
SM values. We quote here the upper bound on the invisible decay rates
of the state discovered at the LHC:\\
\newpage

$(1)~28\% $~~{\rm Ref.}~\cite{Ellis:2013lra};\\[2mm]

$(2)~61\% $~~{\rm Ref.}~\cite{Belanger:2013kya};\\[2mm]

$(3)~69\% $~~{\rm Ref.}~\cite{Espinosa:2012vu, Espinosa:2012im};\\[2mm]

$(4)~30\%$~~{\rm Ref.}~\cite{Giardino:2012ww, Giardino:2012dp}, \\[2mm]

\noindent
consistent with the current data at 95\% confidence level.
In Ref.~\cite{Belanger:2013kya}, it has been pointed out that these
limits can be further improved in the near future with an integrated
luminosity $\mathcal{L} >$ 300 fb$^{-1}$ at $\sqrt{s}$ = 14 TeV at the LHC.
The discovery potential of the $7$ and $8$ TeV LHC in probing the invisible
decaying Higgs has been studied for different final states, where the invisible
Higgs is produced in association with a hard jet (from gluon fusion),
2 jets in the forward direction (from vector boson fusion) or the leptonic
decay of $Z^0$ 
(from associated $Z^0$ production)~\cite{Bai:2011wz, Ghosh:2012ep}.
The invisible decay width of the lightest Higgs boson
has also been investigated 
in MSSM, taking into account the constraints obtained from the
recent data~\cite{Dreiner:2012ex}. 
Recently ATLAS~\cite{ATLAS:2013pma} has looked
for invisible decays of the Higgs with 4.7 fb$^{-1}$ of 7 TeV data
and 13 fb$^{-1}$ of 8 TeV data and has placed limit
on the invisible branching fraction at 95\% confidence level.
They have considered the associated $ZH$ production, with $Z$
decaying leptonically,  and have excluded invisible branching fractions 
greater than 65\%. Being conservative, we consider the
invisible branching fraction to be less than 30\% in this work,
as it is the most constrainedi value.

As mentioned above, since the Higgs and neutralino
sector of NMSSM is quite different from that of  MSSM, conclusions
about the invisible Higgs decay in  MSSM
need to be  reconsidered in the context of
the NMSSM, particularly in relation to the neutralino sector,
as well as the additional possibility of decay into CP odd Higgs bosons. 
In the light of the discovery of the SM-like Higgs boson at the LHC,
considerable work has been done in the context of the Higgs sector of the 
NMSSM~\cite{Forshaw:2007ra, Almarashi:2011te, Belanger:2012tt, 
Ellwanger:2012ke, Gherghetta:2012gb, King:2012tr, Christensen:2013dra}.
These studies  have
scanned various regions of the parameter space, mainly focussing on the 
regions favored by the results from LHC and the flavour physics. 
These studies have also considered the case where the lightest Higgs $h_1$
has a mass of around 100 GeV, and the second lightest scalar $h_2$ is 
identified with the state of mass around $126$ GeV observed at the LHC. 
This is mainly in light of the fact 
that with this assumption the LEP excess~\cite{Barate:2003sz} in the
$e^+e^- \rightarrow Zh \rightarrow Zb\bar{b}$ channel around
$M_{b\bar{b}} \approx 100 GeV$  can be explained together 
with the LHC data. The case with
$h_1$ in the required mass range is also considered for constraining the 
NMSSM parameter space.

One of the  crucial assumptions that go into limiting the  parameter
space of these models is the universality of the gaugino mass parameters
at the grand unified scale~(GUT).
However, the gaugino mass parameters need not be universal at the GUT scale.
If we embed the SM gauge group in a 
grand unified gauge group, the gaugino mass parameters  can be nonuniversal
at the GUT scale, thereby affecting the phenomenology of the 
neutralinos at the weak scale via the renormalization group evolution
of these parameters. This applies to all the 
grand unified theories based on  $SU(5)$, $SO(10)$
and $E_6$ grand unified theories, these being 
the only ones which support the chiral structure of weak interactions 
as observed in nature.

Depending on the gaugino masses at the GUT scale, and hence at the weak scale,
the possibility of massless neutralinos has been considered
in the past~\cite{Gogoladze:2002xp}. 
Such neutralinos could very well be final state particles of the Higgs boson decay.
Neutralinos lighter than half the Higgs mass have not been ruled
out by current data.
In the present work, we consider, among others, the decay of  the lightest 
Higgs boson into lightest neutralinos in low energy supersymmetric models.
This includes the MSSM as well as the NMSSM.
We find that it is not possible to have a massless neutralino
in MSSM, not only with universal gaugino mass parameters $M_1$ and $M_2$
but even with these parameters being nonuniversal at the GUT scale,
except for a higher dimensional
representation of $E_6$. In case of NMSSM, although it is possible to have
massless neutralino with universal gaugino mass parameters at the GUT scale,
it is not possible 
to obtain $m_{h_1}$ = 126 GeV and simultaneously have massless 
neutralinos or $m_{\tilde{\chi}_1^0} \leq m_{h_1}/2$, with universal gaugino 
masses at the GUT scale. We relax the universality assumption on the
gaugino mass parameters, with  $M_1$ and $M_2$ being treated as two 
independent parameters, and consider the question  of light  neutralinos and 
study the decay of the lightest Higgs boson in the context of NMSSM. 
We find that it is possible 
to have large invisible branching ratio for 
$h_1 \rightarrow \tilde{\chi}_1^0 \tilde{\chi}_1^0$. 
The composition of $\tilde{\chi}_1^0$ is important
in determining the invisible branching ratio.
In case of NMSSM, for certain region of the  parameter space
there are additional decay channels. These mainly include the decay of 
$h_1$ to the lightest pseudoscalars, $h_1 \rightarrow a_1a_1,~Z^0a_1$.
These undetected channels will in turn affect the invisible branching 
ratio.

A  very light or massless lightest neutralino which is obtained by considering 
$M_1$ and $M_2$ as independent parameters has to be a bino like, 
since the LEP bound on the chargino mass has set 
lower limits on $M_2$ and $\mu$. Since there is no lower experimental bound on 
this very light neutralino from 
collider experiments, bounds on their properties have been
obtained from other sources. For instance, in 
\cite{Choudhury:1999tn} very light neutralinos together  with R-parity 
violation, consistent with all the experiments, have been proposed as an 
explanation for the KARMEN time anomaly.  Supernova
1987A data has been used to set bounds on the mass
of a nearly pure bino like light neutralino ($m_{\tilde{\chi_1}^0} <$ 200 MeV) 
in the context of 
MSSM~\cite{Dreiner:2003wh}, while gravitino cosmology
with such light neutralinos has been studied in \cite{Dreiner:2011fp}
by taking into account astrophysical and cosmological bounds. 
Moreover a general survey on the bound of the mass of this lightest
neutralino in the context of MSSM with R-parity
conservation has been discussed  in~\cite{Dreiner:2009ic} where
all the collider data along with the contraints from cosmological observations
has been considered.
Overall these studies show that a very light neutralino in the context of
non universal gaugino masses is not ruled out  by current experimental 
observations.

The plan  of the paper is as follows. In Sec.~\ref{sec:gut_boundary}
we consider different
patterns of gaugino masses that arise in  grand unified theories based on
$SU(5)$, $SO(10)$ and $E_6$ gauge groups. We study the existence of a massless
neutralino in these theories with appropriate boundary conditions
as dictated by grand unification.
In Sec.~\ref{sec:mssm}, the decay of the lightest Higgs to neutralinos 
is considered
in the the MSSM case, with the relevant experimental constraints. 
The case of the invisible decay of the lightest Higgs boson for the 
NMSSM is considered in detail in Sec.~\ref{sec:nmssm}. 
The parameter space which 
supports the lightest Higgs $h_1$ in the appropriate mass window 
123-127 GeV is explored. In this Section we also consider the
decay of the lightest Higgs boson to the lightest CP odd Higgs.
Finally,  we summarize our results in Sec.~\ref{sec:conclusions}.
In Appendix A, we briefly summarize some of the details regarding
non-universal gaugino masses in GUTS.
\section{Minimal Supersymmetric Standard Model
with GUT boundary conditions}
\label{sec:gut_boundary}
We begin our analysis with a brief review 
of the existence of a massless or a light neutralino
in the minimal supersymmetric standard model.
We recall that the neutralinos are an admixture of the 
fermionic partners of the two Higgs doublets, $H_1$ and $H_2$,
and the fermionic partners of the neutral gauge bosons. When the 
electroweak symmetry is broken, the physical mass eigenstates 
are obtained from the diagonalization of the neutralino 
mass  matrix~\cite{Bartl:1989ms,Haber:1984rc}  
\begin{eqnarray}
\label{mssmneut}
M_{\mathrm{MSSM}} =
\begin{pmatrix}
M_1 & 0   & - m_Z \sw \cos\beta & \phantom{-}m_Z\sw \sin\beta \\
0   & M_2 & \phantom{-} m_Z \cw \cos\beta  & -m_Z \cw\sin\beta \\
 - m_Z \sw \cos\beta &\phantom{-} m_Z \cw \cos\beta  & 0 & -\mu\\
\phantom{-}m_Z\sw \sin\beta& -m_Z \cw\sin\beta & -\mu & 0
\end{pmatrix},
\end{eqnarray}
\noindent where $M_1$ and $M_2$ are the $U(1)_Y$ and 
the $SU(2)_L$ soft supersymmetry breaking gaugino
mass parameters, $\mu$ is the Higgs(ino) mass parameter, $m_Z$ is the
$Z$ boson mass, $\theta_W$ is the weak mixing angle and 
$\tan\beta=v_2/v_1$ is the ratio of the vacuum expectation values 
of the neutral components of the 
two Higgs doublet fields $H_1$ and $H_2$. We are interested in having a
light neutralino eigenstate of the neutralino mass matrix~(\ref{mssmneut}).
For this purpose we consider the limiting case of the massless neutralino, 
which, at the tree level, arises when the determinant of the 
matrix (\ref{mssmneut}) is zero.  
This leads to the condition~\cite{Gogoladze:2002xp}
\begin{equation}
\mu \left[m_Z^2 \sin 2\beta \left(M_1\cos ^2 \theta_W + M_2 \sin ^2 
\theta_W \right)-M_1M_2 \right] = 0.
\label{det_mssm}
\end{equation}

\noindent 
The solution with $\mu$ = 0 is excluded by the lower bounds on the
chargino mass from the LEP experiments~\cite{ALEPH:2005ab},
which impose the constraint 
\begin{equation}
|\mu |, ~ ~ M_2 \geqslant 100~{\rm GeV}.
\label{mu_bound}
\end{equation}
The other possible solution to (\ref{det_mssm}) can be written as
\begin{equation}
M_1=\frac{M_2 m_Z^2 \sin ^2 \theta_W \sin 2\beta}
{\mu M_2-m_Z^2 \cos ^2 \theta_W \sin 2\beta}.
\label{M1_mssm}
\end{equation}
\noindent 
Therefore,  with fixed values of $\mu, M_2$
and $\tan \beta$, for a massless neutralino,
one must find a value of $M_1$ consistent with
(\ref{M1_mssm}). The condition
(\ref{M1_mssm}) can be expressed in terms of $ r \equiv M_1/M_2$,
so as to check  whether a  massless neutralino is allowed in
the MSSM. In terms of $r$ the condition(\ref{M1_mssm}) can be written as
\begin{equation}
\mu M_2=\frac{m_Z^2}{r} \sin 2\beta(\sin ^2 \theta_W 
+ r \cos ^2 \theta_W),
\label{ratio_mssm}
\end{equation}
\noindent 
which must be satisfied, consistent with the experimental constraints
(\ref{mu_bound}), in order to have a massless neutralino.
 
It is known that the condition
(\ref{ratio_mssm}) is not satisfied in MSSM with universal
gaugino masses at the grand unified scale~\cite{Gogoladze:2002xp}. 
In next section we briefly
recall this and then proceed to study whether this condition can be satisfied 
in MSSM with nonuniversal boundary conditions on the gaugino
mass parameters at the grand unified scale.

\subsection{ Gaugino Masses in Grand Unified Theories}
\label{subsec:universal-gaugino-masses}

In the MSSM, with universal gaugino masses at the grand unified scale, 
usually referred to as mSUGRA, the soft supersymmetry breaking
gaugino mass parameters $M_1, M_2$, and $M_3$  satisfy the  
boundary condition 
\bea
M_1 & = &  M_2 = M_3 = m_{1/2}, \label{gauginogut}
\eea
at the grand unified scale $M_G.$
Furthermore,  the three gauge couplings corresponding to the gauge 
groups $U(1)_Y, SU(2)_L$ and $SU(3)_C$ 
satisfy~($\alpha_i = g_i^2/4\pi, \, i = 1, 2, 3$) 
\bea
\alpha_1 & = & \alpha_2 =  \alpha_3 = \alpha_G,  \label{gaugegut}
\eea
at the GUT scale $M_G$. Using the one-loop 
renormalization group equations~\cite{Martin:1993ft}
for the gaugino masses 
and the gauge couplings this leads to the ratio
\begin{equation}
M_1 : M_2 : M_3 \simeq 1 : 2 : 7.1,
\label{msugra0}
\end{equation}
\noindent
for the soft gaugino masses at the electroweak scale $m_Z$. 
In the following, for definiteness,  we shall consider the
value of $\tan \beta$ = 10.
From Eq.~(\ref{msugra0}), we see that the value of $r$ is 0.5.
Using this in Eq.~(\ref{ratio_mssm}), we conclude that either
$\mu \approx M_2 \approx m_Z$, or $\mu \gg m_Z$ and $M_2 \ll m_Z,$
or $\mu \ll m_Z$ and $M_2 \gg m_Z$. None of these conditions 
are consistent with the LEP constraint~(\ref{mu_bound}).
Thus, a massless neutralino is excluded in the case of 
MSSM with universal gaugino masses at the GUT 
scale.

We recall here that universal soft supersymmetry breaking gaugino
masses are not the only possibility in a grand unified theory. In fact,
non universal boundary conditions for the soft gaugino masses can 
naturally arise in a grand unified supersymmetric theory. 
It is, therefore, important to study whether it is possible to have 
a light neutralino with nonuniversal
boundary conditions at the grand unified scale. To
this end we recall the 
essential features of the boundary conditions on the 
gaugino masses in a grand unified theory. 

\subsection{Nonuniversal Gaugino Masses in  Grand Unified Theories}
\label{subsec:nonuniversal gaugino masses}
We now consider the neutralino masses and mixing in the minimal 
supersymmetric standard model with nonuniversal boundary conditions at 
the GUT scale, which arise in $SU(5)$, $SO(10)$ and $E_6$ grand unified 
theories. As discussed in 
subsection~\ref{subsec:universal-gaugino-masses}, in the 
simplest supersymmetric
model with universal gaugino masses $M_i~(i = 1, 2, 3)$ are
taken to be equal at the grand unified scale. However, in
supersymmetric theories with an underlying grand unified gauge group,
the gaugino masses need not necessarily 
be equal at the GUT scale. 

In  Appendix~\ref{appendix: nonuniversal}, we 
recall the essential features of the embedding of the
SM gauge group in different grand unified gauge groups, namely
$SU(5)$, $SO(10)$ and $E_6$, these being the only ones which support 
the chiral structure of weak interactions as observed in 
nature~\cite{Ramond:1979py}. 
The gaugino mass parameters for the different representations,
that arise in the symmetric
product of the adjoint representations of the respective gauge groups 
are shown in
the Tables~\ref{tab1},~\ref{tab2},~\ref{tab3},~\ref{tab4},~\ref{tab5},
~\ref{tab6},
\ref{tab7},~\ref{tab8},~\ref{tab9},~\ref{tab10} of  
Appendix~\ref{appendix: nonuniversal}.
Using the value of the ratio $r$ at the electroweak
scale from the respective Tables, 
and following the  same procedure as in the case of MSSM 
with universal gaugino masses in the previous subsection, 
we see from  Eq.~(\ref{ratio_mssm}) and 
Tables~\ref{tab1},~\ref{tab2},~\ref{tab3},~\ref{tab4}
that none of the 
representations of $SU(5)$ and $SO(10)$ can have a massless neutralino in
the light of experimental constraints (\ref{mu_bound}). 
We also find that in case of $E_6$, for all the representations
except one, there can be no massless neutralino which satisfies 
the condition (\ref{ratio_mssm}).
Only the higher dimensional  $\bf 2430$
representation  of $E_6$, as shown in Tables~\ref{tab7} and 
\ref{tab8}, with the $\bf 770$
dimensional representation of  $SO(10)$ and a singlet of  $SU(4)',$ 
allows the possibility of  a light neutralino consistent with the 
phenomenological constraint (\ref{mu_bound}). 
We shall not consider this possibility any further in this
paper.

\section{Decay of Higgs to neutralinos in the MSSM}\label{sec:mssm}

In the previous section,  we have seen that in MSSM with universal
gaugino mass parameters at the GUT scale, with  $r = 0.5$
at the weak scale, it is not possible  obtain a massless
neutralino. Since $r \leq 0.04$ for a massless neutralino,
it is not possible to obtain a massless neutralino in a GUT even with 
nonuniversal gaugino masses $M_i$ at the GUT scale. The only possible
exception is 
the higher dimensional representation 
$\bf 2430$ of $E_6$, with $r = 0.02$, and this is not an appealing 
possibility. Thus, in order to obtain a massless neutralino, we must consider
arbitrary gaugino masses in the MSSM.  If the neutralino is
sufficiently light, then the invisible decay  
$h^0 \rightarrow \tilde {\chi}_1^0 \tilde{\chi}_1^0$ will be 
kinematically allowed in MSSM.

Recalling that in the MSSM, the decay width of the lightest Higgs boson to 
a pair of lightest neutralinos can be written as~\cite{Griest:1987qv} 
\begin{equation}\label{higgs_decay1}
\Gamma (h^0 \rightarrow \tilde{\chi}_1^0 \tilde{\chi}_1^0) 
= \frac{G_F m_W^2 m_h}{2 \sqrt{2} \pi}(1- 4m^2_{\tilde{\chi}^0_1}/m_h^2)^{3/2}
\left[(Z_{12} - \tan{\theta_W} Z_{11})
(Z_{13}\sin \alpha + Z_{14}\cos \alpha)\right]^2,
\end{equation}
where $Z_{ij}$ are the elements of the matrix $Z$ which diagonalizes 
the neutralino mass matrix, and $\alpha$ is the mixing angle in the $CP$ 
even Higgs sector. In the decoupling limit, when the mass $m_A$ of the 
pseudoscalar   Higgs boson is large compared to the $Z$ boson mass
$m_Z$, with  $\alpha \rightarrow \beta - \pi/2$,
the decay width (\ref{higgs_decay1}) can be written as~\cite{Dreiner:2012ex}
\begin{equation}\label{higgs_decay2}
\Gamma (h^0 \rightarrow \tilde{\chi}_1^0 \tilde{\chi}_1^0) 
= \frac{G_F m_W^2 m_h}{2 \sqrt{2} \pi}(1- 4m^2_{\tilde{\chi}^0_1}/m_h^2)^{3/2}
\left[(Z_{12} - \tan{\theta_W} Z_{11})
(Z_{14}\sin \beta - Z_{13}\cos \beta)\right]^2.
\end{equation}
\noindent
The composition of the lightest neutralino $\tilde{\chi}_1^0$ 
in terms of the gauginos and Higgsinos can be 
written as~\cite{Gogoladze:2002xp,Bertone:2004pz}
\begin{equation}
 \tilde{\chi}_1^0 = Z_{11}\tilde{B}+Z_{12}\tilde{W}^3+
 Z_{13}\tilde{H}_1^0+Z_{14}\tilde{H}_2^0
\end{equation}
where
\begin{equation}
 Z_{1i}=\left(1,~-\frac{1}{2}\frac{m_Z^2 \sin 2\theta_W \sin 2\beta}
{\mu M_2 - m_Z^2 \cos^2 \theta_W \sin 2\beta},~\frac{m_Z M_2 \sin \theta_W \sin \beta}
 {\mu M_2 - m_Z^2 \cos^2 \theta_W \sin 2\beta},~\frac{m_Z M_2 \sin \theta_W \cos \beta}
 {\mu M_2 - m_Z^2 \cos^2 \theta_W \sin 2\beta}\right).
\end{equation}
\noindent
The invisible decay of the lightest Higgs boson to the lightest
neutralinos, if kinematically allowed,  is mainly constrained by the
$Z$ invisible decay rate.  This invisible decay width
has been measured very precisely by the LEP
experiments~\cite{ALEPH:2005ab} with  
\begin{equation}\label{z_width}
 \Gamma(Z^0\rightarrow \tilde{\chi}_1^0 \tilde{\chi}_1^0) < 3~{\rm MeV}.
\end{equation}
The $Z$ width to a pair of lightest neutralinos 
can be written as \cite{Heinemeyer:2007bw}
\begin{equation}
 \Gamma(Z^0\rightarrow \tilde{\chi}_1^0 \tilde{\chi}_1^0) =
 \frac{G_F m_Z^3}{6\sqrt{2}\pi}(Z_{13}^2-Z_{14}^2)
 \left(1-\frac{4m_{\tilde{\chi}_1^0}^2}{m_{Z^0}^{2}}\right)^{3/2}.
\end{equation}

\begin{figure}[t]
\begin{minipage}[b]{0.45\linewidth}
\centering
\vspace*{0.7cm}
\psfrag{ml}{$\bf{m_{\tilde{e}_L}} $}
\includegraphics[width=6.5cm, height=5cm]{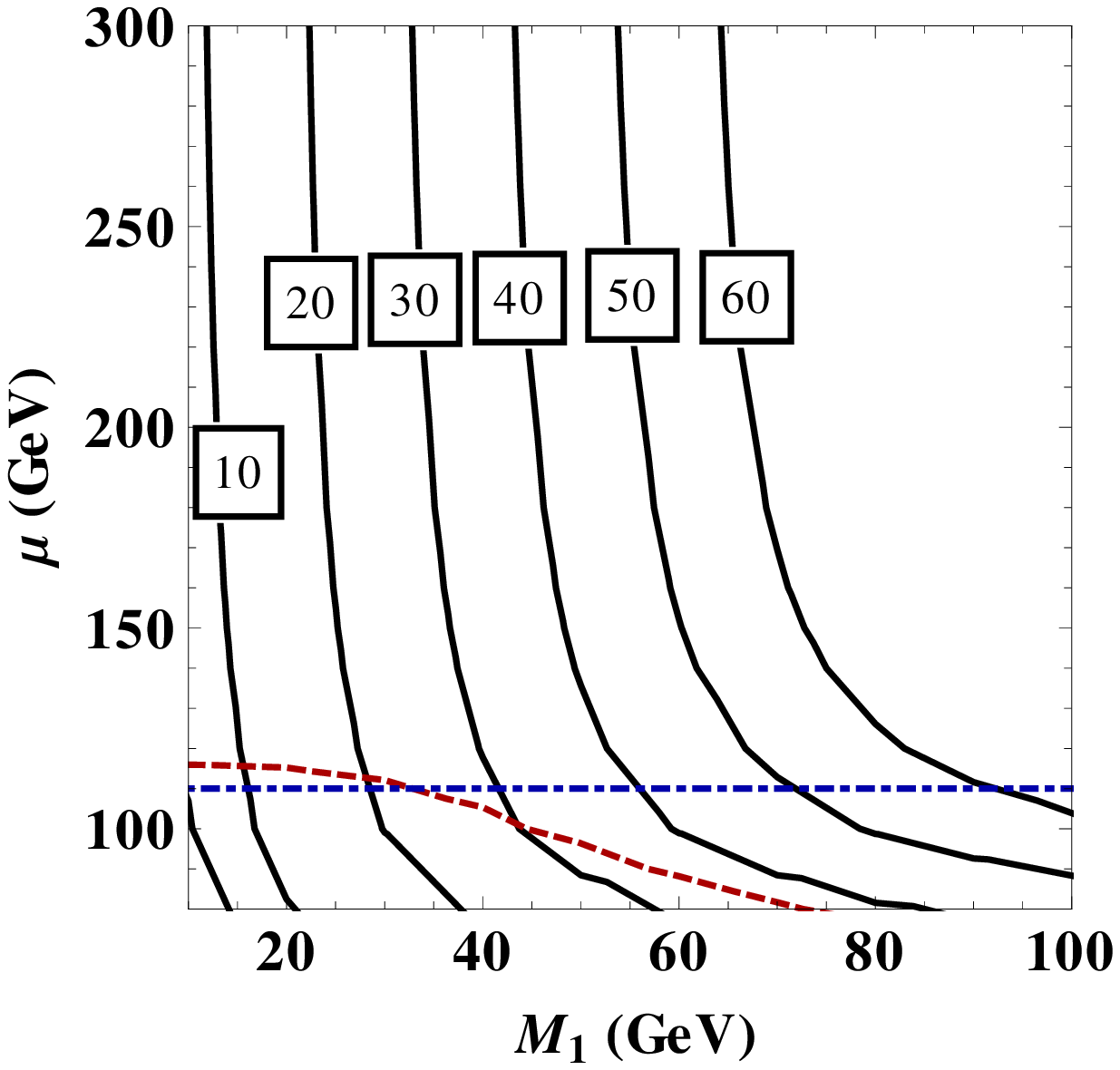}
\caption{The contours of constant lightest neutralino mass in 
MSSM in the $\mu - M_1$ plane for $\tan \beta$ = 10 
and $M_2$ = 200 GeV. }
\label{fig:mssm_ln}
\end{minipage}
\hspace{0.4cm}
\begin{minipage}[b]{0.45\linewidth}
\centering
\psfrag{mr}{$\bf{m_{\tilde{e}_R}} $}
\includegraphics[width=6.5cm, height=5cm]{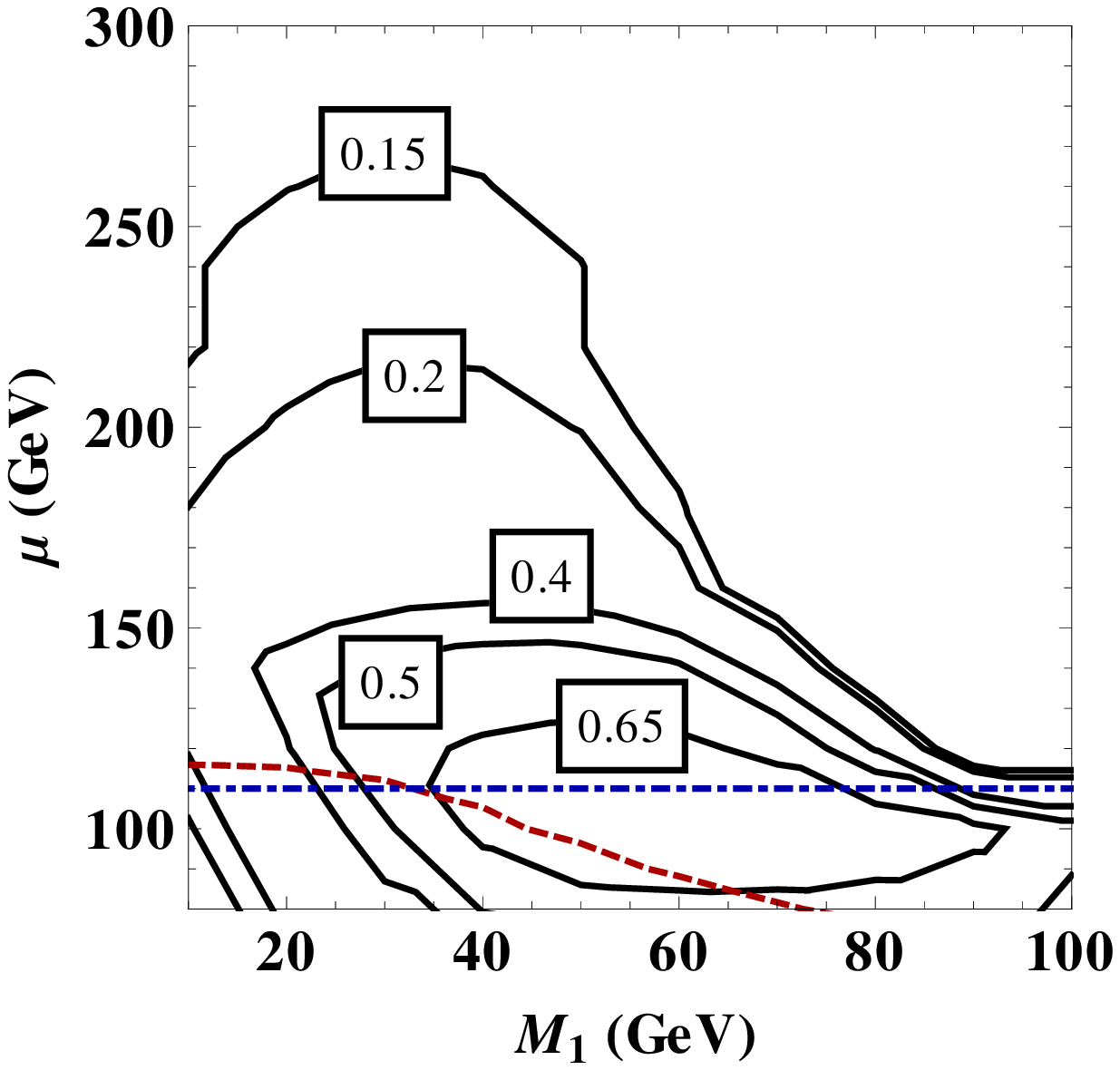}
\caption{The contours of constant  branching ratio of 
$(h\rightarrow \tilde{\chi}_1^0 \tilde{\chi}_1^0)$ 
in MSSM for a fixed value of $\tan \beta$ = 10 and $M_2$ = 200 GeV.}
\label{fig:mssm_br}
\end{minipage}
\end{figure}

For our analysis we have used the program 
CalcHEP~\cite{Belyaev:2012qa}, with $\tan \beta$ = 10.
The trilinear soft supersymmetry breaking coupling $A_t$ has been adjusted
in order to obtain a lightest Higgs boson  of mass $\approx$ 126 GeV. 
The gluino mass is taken to be 
1400 GeV~\cite{ATLAS:2012sma}, and the squarks are 
assumed to have a mass above 1 TeV~\cite{CMS:2012yua},
thereby respecting the current  experimental bounds. We have presented our 
results for a fixed value of $M_2$, with the parameters $\mu$ and
$M_1$ being varied. Since the results don't change significantly  as a function
of $M_2$, only a particular value of $M_2$ is considered. 
In Fig.~\ref{fig:mssm_ln} we  show 
the contour plots of the constant lightest neutralino mass in MSSM,
and in Fig.~\ref{fig:mssm_br} the corresponding  contours of constant invisible 
branching ratio of the lightest Higgs boson.
In our calculations we have imposed the constraint of the 
lightest  chargino mass bound 
$m_{\tilde{\chi}^+} > 94$ GeV from the LEP experiments as well as 
the bound from invisible
$Z^0$ decay width coming from $Z^0$ decay into neutralinos. 
Our results agree with those of  Ref.~\cite{Dreiner:2012ex}.
This sets the stage for our analysis of the invisible decay of the
lightest Higgs boson in the NMSSM, which we carry out in the next section.

\section{Decay of the lightest Higgs to neutralinos and pseudoscalars
in the NMSSM}\label{sec:nmssm}
The NMSSM is characterized by the presence of the
gauge singlet superfields $S$ in addition to the two Higgs doublets
$H_1$ and $H_2$ of the minimal supersymmetric standard model. 
The Higgs(ino) mass term $\mu H_1 H_2$
in the superpotential of the MSSM is replaced by the 
trilinear coupling $\lambda S H_1 H_2$
where  $\lambda$ is a dimensionless 
coupling~\cite{Fayet:1974pd, Ellis:1988er, Drees:1988fc, Pandita:1993hx,
Pandita:1993tg, Ellwanger:1993hn, Elliott:1993uc}. 
In addition there is
also a trilinear self coupling of the singlet, namely $S^3.$
The part of the superpotential involving only 
the Higgs superfields has the form
\begin{equation}\label{nmssm_sp}
W_{NMSSM}=\lambda SH_1 H_2 -\frac{\kappa}{3}S^3.
\end{equation}
After the electroweak symmetry breaking, the vacuum expectation value 
(VEV) of the singlet field,  $<S> \equiv x$ generates an effective 
$\mu$ parameter,
$\mu_{eff} = \lambda x$, which is naturally of order of the 
electroweak scale, thus providing a solution to the $\mu$ problem 
of the MSSM.  Thus,  compared to the two independent parameters
in the Higgs sector of the MSSM at tree
level ($\tan \beta, M_A$), the Higgs sector of NMSSM is described
by six parameters $\mu_{eff}$, $\lambda$, $\kappa$, $\tan \beta$,
$A_{\lambda}$ and $A_{\kappa}$, where $A_{\lambda}$ and $A_{\kappa}$
are the trilinear supersymmetry breaking couplings.

Due to the addition of the  singlet, the neutralino
mass matrix in NMSSM is a $5 \times 5$ matrix, which 
in the bino, wino, Higgsino and singlino basis can be  written 
as~\cite{Pandita:1994ms, Pandita:1994vw, Choi:2004zx} 
\begin{eqnarray}
\label{nmssmneut}
M_{\mathrm{NMSSM}} = \begin{pmatrix}
M_1 & 0   & - m_Z \sw \cos\beta & \phantom{-}m_Z\sw \sin\beta & 0 \\
0   & M_2 & \phantom{-} m_Z \cw \cos\beta  & -m_Z \cw\sin\beta & 0 \\
- m_Z \sw \cos\beta &\phantom{-} m_Z \cw \cos\beta  & 0 & -\mu_{eff}
& -\lambda v_2\\
\phantom{-}m_Z\sw \sin\beta& -m_Z \cw\sin\beta & -\mu_{eff} & 0 & -\lambda v_1\\
0 & 0 &  -\lambda v_2 & -\lambda v_1 & 2 \kappa x
\end{pmatrix}.
\end{eqnarray}
\noindent
The neutralino sector in this case 
is described by six parameters, $\mu_{eff}$, $M_1$, $M_2$, 
$\tan \beta$, $\lambda$ and $\kappa$.
For a massless neutralino the
determinant of the mass matrix  (\ref{nmssmneut})
should be zero, which leads to~\cite{Gogoladze:2002xp}
\begin{eqnarray}
2 \kappa x \mu_{eff} (\Delta_{0} \sin 2 \beta - \mu_{eff} M_1 M_2 )
+ \lambda^2 v^2\left[\Delta_{0}
-\mu_{eff} M_1 M_2 \sin 2 \beta \right] = 0,
\label{det_nmssm}
\end{eqnarray}
\noindent
where $\Delta_{0} = m_{Z}^{2} (M_1 \cos^2 \theta_W + M_2 \sin^2 \theta_W)$.
Eq.~(\ref{det_nmssm}) in turn leads to the following condition 
\begin{equation}\label{nmssm_mn}
\kappa = \frac{\lambda}{2}\left(\frac{\lambda v}{\mu_{eff}}\right)^2
\frac{\Delta_{0}
-\mu_{eff} M_1M_2 \sin 2\beta}{\mu_{eff} M_1 M_2 -\Delta_{0}\sin 2\beta},
\end{equation}
\noindent
for a massless neutralino in the NMSSM.
The composition of the lightest neutralino $\tilde{\chi}_1^0$ 
in terms of the gauginos, Higgsinos and the singlino is in 
turn given by
\begin{equation}
\tilde{\chi}_1^0 = Z'_{11}\tilde{B}+Z'_{12}\tilde{W}^3+
 Z'_{13}\tilde{H}_1^0+Z'_{14}\tilde{H}_2^0 + Z'_{15} S,
\end{equation}
where
\begin{eqnarray}
Z'_{1i}&=&\left(-\frac{\lambda v m_Z \cos 2\beta \sin \theta_W M_2}
 {\Delta_1},
 ~~\frac{\lambda v m_Z \cos 2\beta \cos\theta_W M_1}
 {\Delta_1}, \right. \nonumber \\
 &&\left. ~~\frac{v(\sin \beta \Delta_0 -\mu_{eff} M_1 M_2 \cos \beta)}
 {x \Delta_1}
 ,~~\frac{v(\cos \beta \Delta_0 -\mu_{eff} M_1 M_2 \sin \beta)}
 {x \Delta_1 },~~1
\right)
\end{eqnarray}
and $\Delta_1 = \mu_{eff} M_1 M_2 - \Delta_{0} \sin 2 \beta$. 
Here $Z'$ is the matrix which diagonalizes the $5 \times 5$
neutralino mass matrix of the NMSSM. As in the case of MSSM,
we  have assumed $CP$ conservation in the neutralino sector in our analysis.

We have performed our analysis for the NMSSM with the set of  
relevant parameters varied in the following ranges:
\begin{enumerate}
\item 4 $\leq\tan \beta \leq$ 11, 
$100~{\rm GeV} \leq \mu_{eff} \leq 200~{\rm GeV}$,
\item $0.55 \leq \lambda \leq 0.7$, $0.33 \leq \kappa \leq 0.8$,
\item $-10$ GeV $\leq A_{\kappa} \leq$ 10 GeV, 
500 GeV $\leq A_{\lambda} \leq$ 1000 GeV.
\end{enumerate}
This range is considered, because we are mainly interested in the
region where the lightest CP even Higgs ($h_1$) of the the NMSSM will lead 
to a SM like Higgs in the mass range $124~{\rm GeV}\leq m_{h_1}\leq 127$ GeV. 
We have restricted ourselves to small values of $\tan \beta$, since
it is  difficult to get a SM like lightest Higgs in the mass window
of 124 - 127 GeV with larger values of $\tan \beta$. The
range for $\lambda$ and $\kappa$ are chosen by imposing
the theoretical constraint that 
there are no charge and color breaking global minima of 
the scalar potential and that a Landau
pole does not develop below the GUT scale.  We are
interested mainly in relatively large values of $\lambda$, so as to
increase the tree level mass of the CP-even Higgs boson,
leading naturally to a SM like Higgs bosons. This in turn implies
a large doublet singlet mixing in the Higgs sector.
The lightest Higgs boson with mass 
$\approx$ 126 GeV can also be achieved in NMSSM, as in  MSSM through loop level
corrections coming from stop,  with large values of $A_t$. 
In this case $\lambda$ can be small 
($\lambda \approx$ 0.1), typically preferred for negative values of 
$A_\kappa$. Here we have considered the former case, where the Higgs mass is 
obtained naturally at tree level. Since we are mainly interested
in large $\lambda$, the other NMSSM parameters are considered
accordingly so as to satisfy the constraints from precision
electroweak measurements, see Ref.~\cite{Cao:2008un}. 
In addition we have also taken into account the latest
experimental constraints from the LHC on the gluino
and other sparticle masses. The gluino mass is chosen
above 1400 GeV, and the squark masses are set to 1 TeV or more,
as in the  MSSM analysis.
Additional constraints from  $B$ physics and
the anomalous magnetic moment of muon are taken into account 
using CalcHEP, which has inbuilt NMSSMTools 
package~\cite{Ellwanger:2004xm, Ellwanger:2005dv}.
In Table~\ref{tab:nmssmparam} we summarize 
the values of the various input parameters used for our analysis.
\begin{table}[htb]
\renewcommand{\arraystretch}{1.0}
\begin{center}
\vspace{0.5cm}
\begin{tabular}{|c|c|c|c|}
\hline
$\tan\beta$ = 10 &$\mu_{eff}$ = 130 GeV &$A_\lambda$ = 880 GeV 
&$A_\kappa$ = 10 GeV \\
\hline
$M_3$ = 1402 GeV    &$A_t$= 2800 GeV &$A_b$= 2800 GeV &$A_{\tau}$= 1000 GeV \\
\hline 
\end{tabular}
\end{center}
\vspace{-0.5cm}
\caption{Input parameters for the  NMSSM}
\renewcommand{\arraystretch}{1.0}
\label{tab:nmssmparam}
\end{table}
\noindent
Considering the relation between $M_1, M_2$ and $M_3,$ choosing
the $SU(3)_C$ gaugino mass parameter $M_3$ = 1402 GeV, with 
the remaining two soft SUSY
breaking gaugino parameters having values  $M_1$ = 197 GeV
and $M_2$ = 395 GeV, respectively. With this, and using (\ref{nmssm_mn}),
{\it we find that it is not possible to get a 
massless neutralino in the  NMSSM, 
with $m_{h_1} \approx 126$ GeV.}
We arrive at this conclusion by taking into
account the experimental constraint (\ref{mu_bound}).
This result holds in the  entire parameter space
considered in our analyses.
If the condition, $m_{h_1} \approx 126$ GeV, is relaxed with
the mass of the next to lightest CP even Higgs $m_{h_2}$
to be in the mass range 124 - 127 GeV, then
it is possible to obtain a massless neutralino. 
We do not consider this possibility  here. 
{\it Thus, for NMSSM, in the  region of the parameter space
considered by us, universal boundary conditions on the gaugino masses 
at the GUT scale cannot
lead to a decay for $h_1\rightarrow \tilde{\chi}_1^0\tilde{\chi}_1^0$,
since $m_{\tilde{\chi}_1^0}\geqslant m_{h_1}/2$ in this region, i.e. the 
decay is not kinematically possible.}
This can be seen from Table~\ref{tab:nmssmpoints}, 
where we present  the values of $m_{h_1}$, $m_{\tilde{\chi}_1^0}$
and $m_{a_1}$, for different  combinations of $\lambda$ and $\kappa$.
The other parameters are fixed, with the values considered
in Table~\ref{tab:nmssmparam}. It can be easily seen from
Table~\ref{tab:nmssmpoints}, that for the mass of
$m_{h_1}$  around 126 GeV, the lightest neutralino mass varies
in the range 80 - 90 GeV. Therefore, the invisible decay to the
lightest neutralinos is  not  kinematically allowed. 
This  result is
also true, when $\lambda$ is small, as discussed before, for the 
case where the lighest Higgs achieves mass through loop corrections.
We have found $m_{\tilde{\chi}_1^0}\geqslant m_{h_1}/2$
by scanning the entire parameter $(\lambda, \kappa)$
space with  $0.001 \leq \lambda \leq 0.7$
and $0.001 \leq \kappa \leq 0.8$. The dependence of 
our results on the other
input parameters which were fixed  for this analysis will be discussed 
in the following.

\begin{table}[htb]
\begin{center}
\begin{tabular}{|c|c|l|c|c|c|c|c|c|c|c|c|c|c|c|c|} \hline
&\multicolumn{3}{|c|}{}&\multicolumn{3}{|c|}{}
&\multicolumn{3}{|c|}{} 
&\multicolumn{3}{|c|}{} &\multicolumn{3}{|c|}{}\\
 &  \multicolumn{3}{|c|}{$\kappa$ = 0.33} 
 &  \multicolumn{3}{|c|}{$\kappa$ = 0.43} 
 &  \multicolumn{3}{|c|}{$\kappa$ = 0.53}
 &  \multicolumn{3}{|c|}{$\kappa$ = 0.63} 
 &  \multicolumn{3}{|c|}{$\kappa$ = 0.73}\\[2mm]\cline{2-16}
&&&&&&&&&&&&&&& \\
$\lambda$ & $m_{h_1}$ & $m_{\tilde{\chi}_1^0}$ &$m_{a_1}$
 & $m_{h_1}$ & $m_{\tilde{\chi}_1^0}$ &$m_{a_1}$
 & $m_{h_1}$ & $m_{\tilde{\chi}_1^0}$ &$m_{a_1}$
 & $m_{h_1}$ & $m_{\tilde{\chi}_1^0}$ &$m_{a_1}$
 & $m_{h_1}$ & $m_{\tilde{\chi}_1^0}$ &$m_{a_1}$ \\[2mm] \hline\hline
0.55 &113 &73.7 &46.1 &122 &82.9 &51.4 &125 &88.7 &55.7
&125.6 &92.5 &59.2 &126 &95.1 &62.2 \\[2mm]
0.58 &108.3 &69.5 &50 &120 &79.1 &55.9 &124 &85.5  &60.9
&125 &89.8 &65 &126 &92.8 &68.6\\[2mm]
0.61  &102.6 &65.4 &53.5  &117.9 &75.4 &60 &123 &82.2 &65.5
&124.9 &86.9 &70.1 &125.8 &90.4 &74.2\\[2mm]
0.64 &96.4 &61.4 &56.7 &114.9 &71.6 &63.8 &121.7 &78.8 &69.7
&124.3 &83.9 &74.8 &125.5 &87.7 &79.2 \\[2mm]
0.67 &89.8 &57.5 &59.7 &111.1 &67.8 &67.2 &119.9 &75.3 &73.5
 &123.4 &80.9 &79.1 &125 &84.9 &83.9   \\[2mm]
0.7   &82.7 &53.9 &62.4 &106.5 &64.1 &70.3 &117.7 &71.9 &77.1
 &122 &77.7 &82.9 &124 &82.1  &88.2 \\[2mm]
\hline
\end{tabular}
\end{center}
\caption{The mass of the lightest CP even Higgs $h_1$, lightest
neutralino $\tilde{\chi}_1^0$ and the lightest CP odd pseudoscalar
Higgs $a_1$, in the NMSSM with universal gaugino masses at the GUT scale,
for the parameter space considered in Table~\ref{tab:nmssmparam}
and with $M_1$ = 197 GeV, $M_2$ = 395 GeV.}
\label{tab:nmssmpoints}
\end{table}

\begin{figure}[htb]
\includegraphics[width=6.5cm, height=5cm]{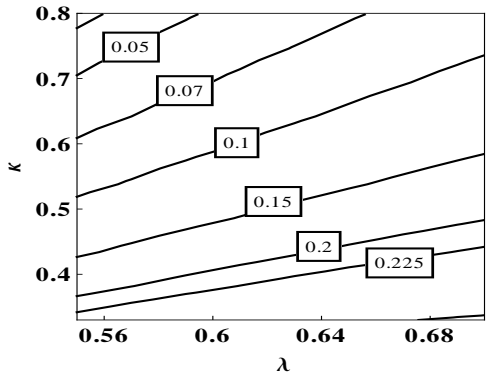}
\caption{Contours of constant singlino composition
$|Z'_{15}|^2$ in the $\kappa - \lambda$ plane for NMSSM, with universal 
gaugino masses at the GUT scale with fixed values of
$M_1$ = 197 GeV, $M_2$ = 395 GeV, and
the other input parameters as given in Table~\ref{tab:nmssmparam}.}
\label{fig:nmssm_sing3}
\end{figure}

It may be noted that in the case of the NMSSM,  
the lightest neutralino has a singlino component
along with the gaugino and Higgsino components.
We have analysed the singlino component of $\tilde{\chi}_1^0$
in the parameter space $\lambda$ and $\kappa$, with the other parameters
fixed at the values as in Table~\ref{tab:nmssmparam}, and with
$M_1$ = 197 GeV, $M_2$ = 395 GeV, respectively.
The  gaugino plus Higgsino, and the singlino components
are respectively given by $Z_{11}^{'2}+Z_{12}^{'2}$, 
$Z_{13}^{'2}+Z_{14}^{'2},$
and $Z_{15}^{'2}$. The decay width of $h_1$ to the lightest neutralino
in NMSSM can be written as~\cite{Franke:1995tc, Franke:1995tf}: 
\begin{equation}\label{nmssm_higgs1}
\Gamma (h_1 \rightarrow \chi_1^0 \chi_1^0) = \frac{m_{h_1}}{16 \pi}(1- 4m^2_{\chi^0_1}/{m_{s_1}^2})^{3/2}
Q^{{\prime}{\prime}{L^2}}_{111}
\end{equation}

\begin{eqnarray}\label{nmssm_higgs2}
Q^{{\prime}{\prime}}_{111} &=& \left[\frac{g}{c_W} Z'_{12}((U_{11}^s \cos \beta +U_{12}^s \sin \beta) 
Z'_{13} + (U_{11}^s \sin \beta - U_{12}^s \cos \beta) Z'_{14}) \right. \nonumber \\ 
&& \left. +\sqrt{2} \lambda Z'_{15}(U_{11}^s \cos \beta 
+ U_{12}^s \sin \beta) Z'_{14} - (U_{11}^s \sin \beta 
- U_{12}^s \cos \beta) Z'_{13})\right] \nonumber \\
&&-2 \sqrt{2} \kappa U_{13}^s |Z'_{15}|^2
\end{eqnarray}
where $U^s$ is the matrix that diagonalizes the $3 \times 3$ scalar
Higgs mass matrix of the NMSSM.  It  is clear from
Eq.~(\ref{nmssm_higgs2}), that as the singlino
contribution appears with a negative  sign in the decay width, 
the invisible decay width of $h_1$ would decrease, as
the singlino composition increases. 
Nevertheless no simple explanation is available, since in practice either
sign solutions for the singlino matrix element are found.
We show in Fig.~\ref{fig:nmssm_sing3}, the 
contours of constant singlino component, 
in the case with universal gaugino masses 
at the GUT scale, where we see that there is a significant singlino
component in the lightest neutralino.
For lower values of $M_1,$
the lightest neutralino has a dominant gaugino component.
Since in this case $M_1$ is 
around 180 GeV, due to the constraint on the  gluino mass, the gaugino
and Higgsino components  decrease, with the neutralino being dominantly a
singlino.

\begin{table}[htb]
\begin{center}
\begin{tabular}{|c|c|l|c|c|c|c|c|c|c|c|c|c|c|c|c|} \hline
&\multicolumn{15}{|c|}{}\\
&  \multicolumn{15}{|c|}{$M_1$ (GeV)} \\[2mm]\cline{2-16}
&\multicolumn{3}{|c|}{}&\multicolumn{3}{|c|}{}
&\multicolumn{3}{|c|}{} 
&\multicolumn{3}{|c|}{} &\multicolumn{3}{|c|}{}\\
 &  \multicolumn{3}{|c|}{$\kappa$ = 0.33} 
 &  \multicolumn{3}{|c|}{$\kappa$ = 0.43} 
 &  \multicolumn{3}{|c|}{$\kappa$ = 0.53}
 &  \multicolumn{3}{|c|}{$\kappa$ = 0.63} 
 &  \multicolumn{3}{|c|}{$\kappa$ = 0.73}\\[2mm]\cline{2-16}
&&&&&&&&&&&&&&& \\
$\lambda$ & 20 & 50 &80
 & 20 & 50 &80 & 20 & 50 &80
 & 20 & 50 &80 & 20 & 50 &80 \\[2mm] \hline\hline
0.55 &4.67 &0.33 &0.02 &3.08 &0.21 &0.04 &2.43 &0.13 &0.03
&2.1 &0.10 &0.01 &1.89 &0.08 &0.01 \\[2mm]
0.58 &6.33 &0.14 &0.34 &3.77 &0.16 &0.04 &2.83 &0.13 &0.05
&2.36 &0.09 &0.05 &2.09 &0.06 &0.04\\[2mm]
0.61 &9.13  &$\approx 10^{-4}$ &0.13 &4.80
&$\approx 10^{-4}$ &0.02 &3.37 &$\approx 10^{-4}$ &0.05
&2.71 &$\approx 10^{-3}$ &0.06 &2.34 &$\approx 10^{-2}$& 0.06 \\[2mm]
0.64 &14.3 &0.05 &0.11 &6.40 &0.02 &0.31 &4.15 &0.03 &0.02
& 3.17 &0.07 & 0.05 &2.65 &0.11 &0.06\\[2mm]
0.67  &25.8 &0.44 &0.05 &9.01 &0.42 &0.08 &5.29 &0.40 &0.03
&3.83 &0.33 &0.04 &3.09 &0.24 &0.05  \\[2mm]
0.7 &68.3 &0.68 & $\approx 10^{-5}$ &13.65 &0.79 &0.02
& 70.57 &0.73 &0.06 &4.76 &0.51 &0.01 &3.67 &0.34 &0.03 \\[2mm]
\hline
\end{tabular}
\end{center}
\caption{The relic density of the lightest neutralino, for different values
of $M_1$, for the parameter space considered in Table~\ref{tab:nmssmparam}
and with $M_2$ = 200 GeV.}
\label{tab:relicdensity}
\end{table}

Here it is  important to consider the possibility that our parameter
choice could lead to over closure of the universe. We use 
MicroOmegas~\cite{Belanger:2005kh, Belanger:2006is}
implemented in NMSSMTools to compute the dark matter relic density of the 
lightest neutralino, $\tilde{\chi}_1^0$.
We show in Table~\ref{tab:relicdensity}, the corresponding relic 
density for different values of $M_1$, in the $\kappa-\lambda$
parameter space. The measurements from
WMAP has constrained the relic density of dark matter~\cite{Bennett:2003ca},
i.e. (0.0925 $< \Omega h^2 <$ 0.1287). It can be seen from the Table,
that the relic density constrains most of the ($\lambda - \kappa$)
parameter space, depending on the value of $M_1$. When the lightest 
neutralino is mostly a bino, due to a small value of $M_1$,  the relic
density is sufficiently large at smaller values of  $\kappa$ and
larger values of $\lambda$. This has to do with the dependence of neutralino
mass on $\lambda$ and $\kappa$, which will be discussed later. The relic 
density mostly constrains smaller values of $M_1 <$ 50 GeV, and as will
be seen later this region is  disfavored by the Higgs invisible branching
ratio. Thus we see that our choice does not come in conflict with the 
cosmological relic density constraint.

\begin{figure}[htb]
\begin{minipage}[b]{0.45\linewidth}
\centering
\vspace*{0.7cm}
\includegraphics[width=7cm, height=5cm]{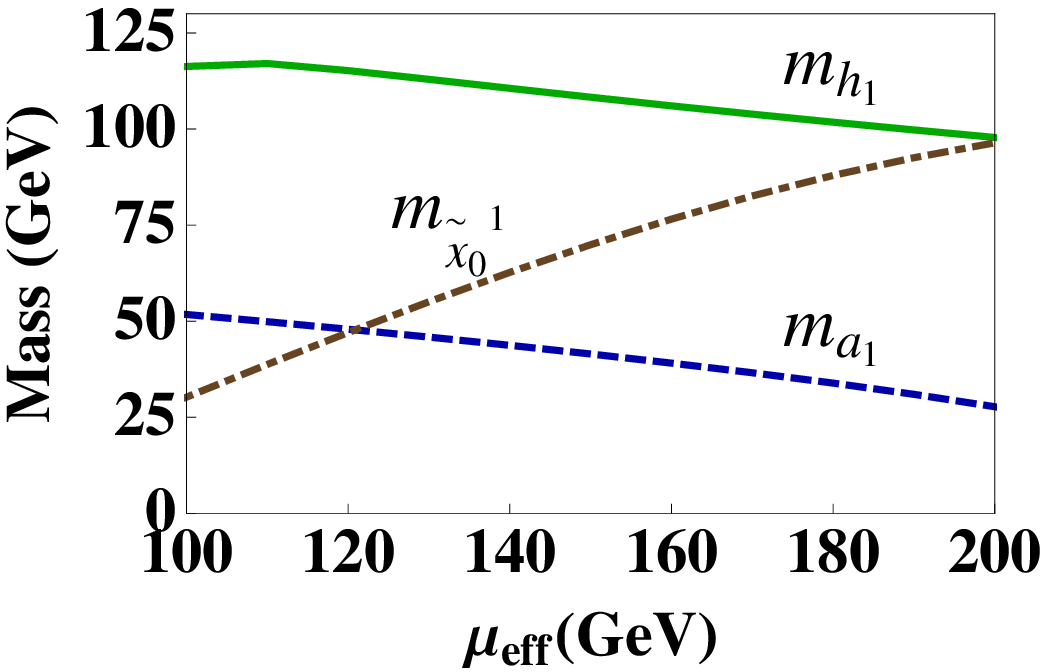}
\caption{Dependence of $m_{h_1}$ [green-solid], $m_{a_1}$ [blue-dashed]
and $m_{\tilde{\chi}_1^0}$ [brown-dot-dashed] on $\mu_{eff}$,
for $M_1$ = 120 GeV, $M_2$ = 200 GeV, $\lambda$ = 0.55 and $\kappa$ = 0.33 with
the other input parameters fixed to the values given in 
Table~\ref{tab:nmssmparam}. (colours in on-line version)}
\label{fig:mueff}
\end{minipage}
\hspace{0.4cm}
\begin{minipage}[b]{0.45\linewidth}
\centering
\includegraphics[width=7cm, height=5cm]{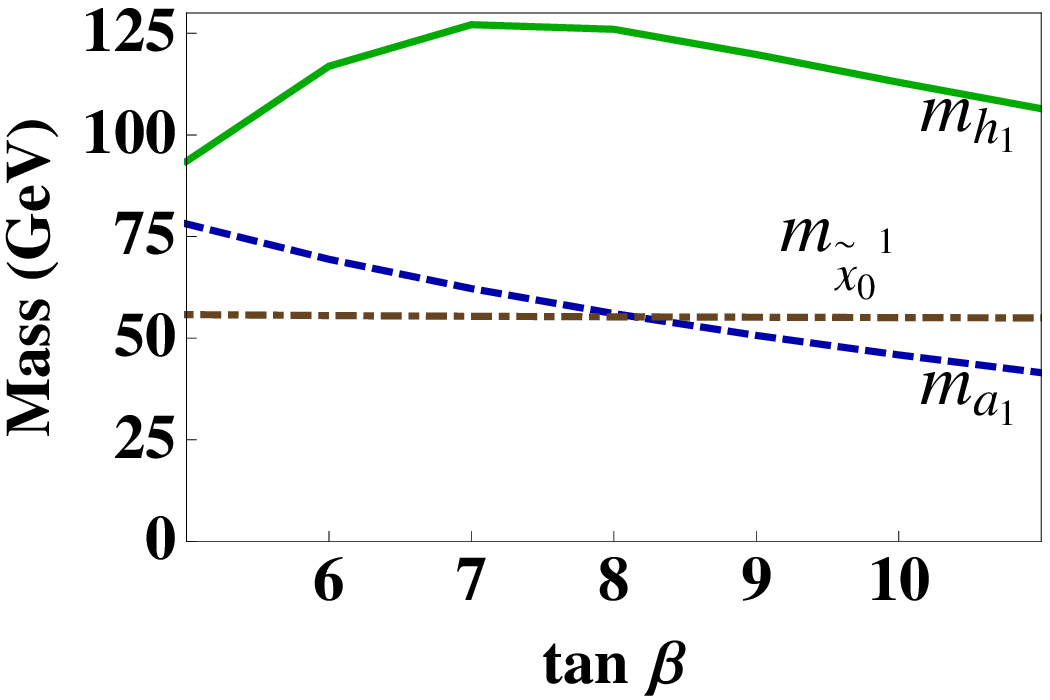}
\caption{Dependence of $m_{h_1}$ [green-solid], $m_{a_1}$ [blue-dashed]
and $m_{\tilde{\chi}_1^0}$ [brown-dot-dashed] on $\tan \beta$
for $M_1$ = 120 GeV, $M_2$ = 200 GeV, $\lambda$ = 0.55 and $\kappa$ = 0.33
with the other input parameters fixed to the values given in 
Table~\ref{tab:nmssmparam}.(colours in on-line version)}
\label{fig:tanbeta}
\end{minipage}
\end{figure} 

With the universal gaugino masses at the GUT scale,
the Higgs invisible decay to the lightest neutralinos
is kinematically not allowed in the NMSSM. We, therefore,
use $M_1$ and $M_2$ as two independent parameters.
Before proceeding further we would like to comment on the
dependence of our results on the various input parameters considered
in  our analysis. For this we consider the dependence of the mass of the
lightest CP even Higgs $h_1$, the lightest pseudoscalar Higgs $a_1$
and the lightest neutralino $\tilde{\chi}_1^0$ on different NMSSM parameters,
$\mu_{eff}$, $\lambda$, $\kappa$, $\tan \beta$, $A_\lambda$ and
$A_\kappa$. We consider the mass of the lightest pseudoscalar Higgs 
because for certain regions of the parameter space it is rather 
light, in fact $a_1$
can  be lighter than $h_1$. This could
lead to additional decay channels for the lightest CP even Higgs, mainly the 
channel $h_1\rightarrow a_1 a_1,$ and $h_1\rightarrow a_1 Z^0$. 
In the observed mass window of the Higgs, the
decay to $b\bar{b}$ is dominant, but with the additional decay
channel $h_1\rightarrow a_1 a_1$ and 
$a_1 \rightarrow b\bar{b}, \tau \bar{\tau}, \mu \bar{\mu}, 
\tilde{\chi}_1^0 \tilde{\chi}_1^0$, depending
on the mass of the lightest pseudoscalar, 
the branching fraction  $h_1 \rightarrow b\bar{b}$ can be 
significantly reduced. It may be emphasized that 
the LHC sensitivity in case of Higgs decay to
light pseudoscalars depend on the decay mode of the pseudoscalars. 
For the parameter space considered in our analyses, $a_1$
mainly decays to $b\bar{b}$. At the LHC, this channel  will be dominated
by a large QCD background. The $b\bar{b}$ channel in the Higgs decay
has been searched for at the LHC, and indicates weak SM Higgs signal
of around 1-2 $\sigma$.
This particular decay channel of $h_1$ decaying to pseudoscalar $a_1$
pairs has also been discussed 
in~\cite{Dermisek:2005gg, Dermisek:2008uu, Dermisek:2010mg, 
Calderini:2012ar, Englert:2012wf, Cerdeno:2013cz}.
\begin{figure}[htb]
\begin{minipage}[b]{0.45\linewidth}
\centering
\vspace*{0.7cm}
\includegraphics[width=7cm, height=5cm]{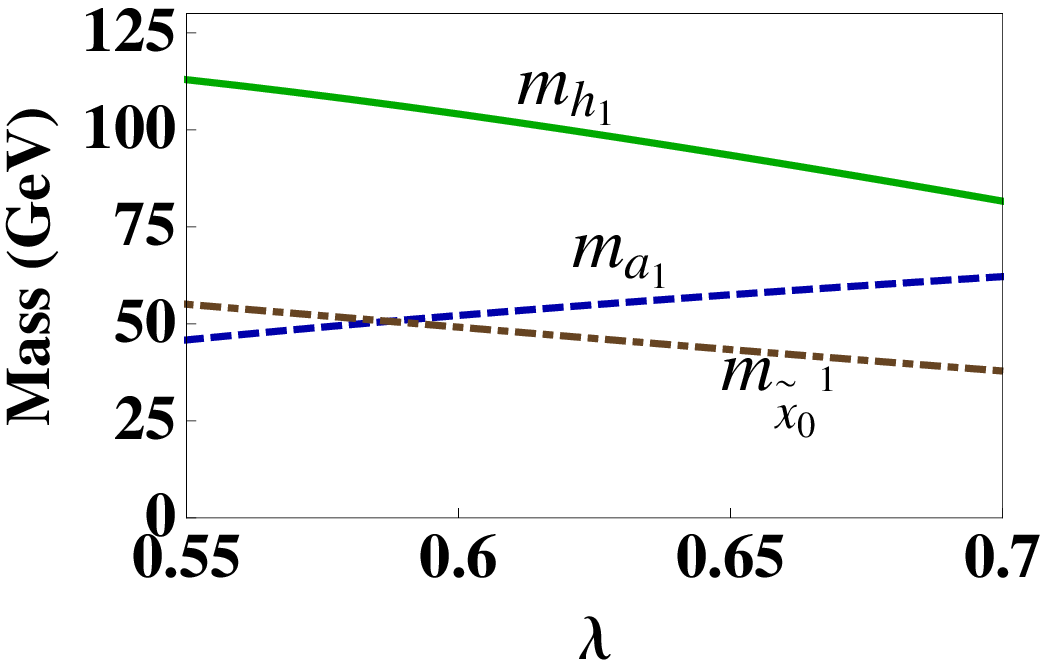}
\caption{Dependence of $m_{h_1}$ [green-solid], $m_{a_1}$ [blue-dashed]
and $m_{\tilde{\chi}_1^0}$ [brown-dot-dashed] on $\lambda$
for $M_1$ = 120 GeV, $M_2$ = 200 GeV and $\kappa$ = 0.33 with
the other input parameters fixed to the values given in 
Table~\ref{tab:nmssmparam}.(colours in on-line version)}
\label{fig:lambda}
\end{minipage}
\hspace{0.4cm}
\begin{minipage}[b]{0.45\linewidth}
\centering
\includegraphics[width=7cm, height=5cm]{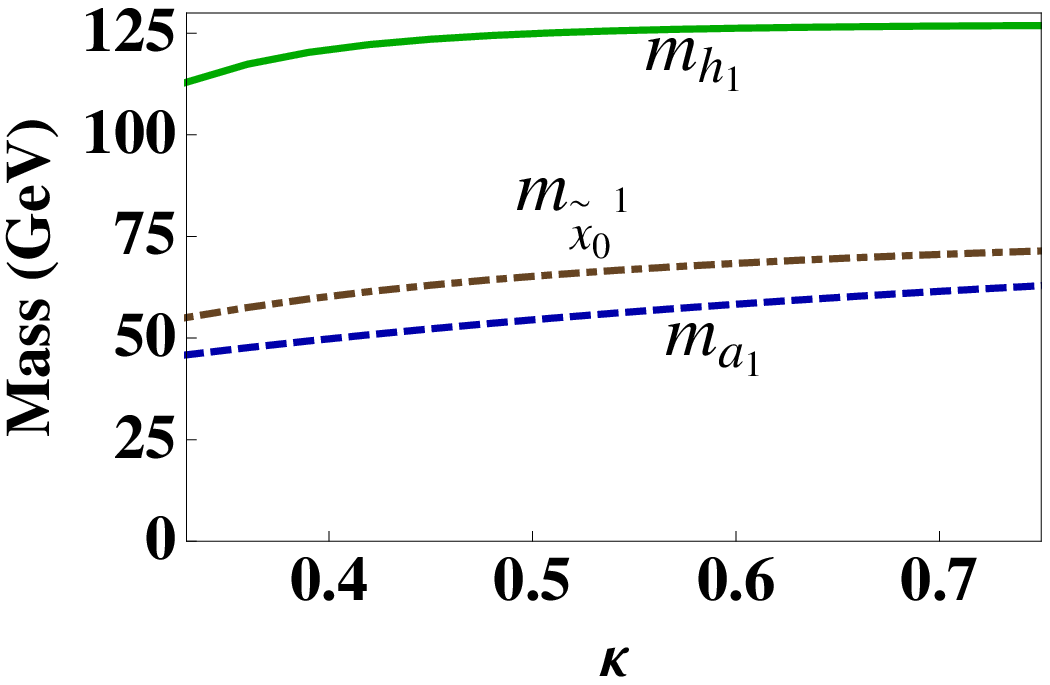}
\caption{Dependence of $m_{h_1}$ [green-solid], $m_{a_1}$ [blue-dashed]
and $m_{\tilde{\chi}_1^0}$ [brown-dot-dashed] on $\kappa$
for $M_1$ = 120 GeV, $M_2$ = 200 GeV and $\lambda$ = 0.55 with
the other input parameters fixed to the values given in 
Table~\ref{tab:nmssmparam}. (colours in on-line version)}
\label{fig:kappa}
\end{minipage}
\end{figure} 

Most of the studies in the context of the lightest pseudoscalar
have been carried out in the light of  the LEP constraints on the Higgs mass, 
$m_h > 114$ GeV, along with the LEP excess
for a lighter Higgs around 100 GeV, through $Z^0h$ production, 
where $h$ decays primarily
to $b$ quarks. It has been  concluded that if 
in the NMSSM, the Higgs boson decays mainly into $a_1$ pairs,
and with $m_{a_1} < 2m_b$,
then the LEP constraints can be evaded. It will be possible to 
have a lighter Higgs of mass less than 105 GeV,
satisfying all precision electroweak results. This is often
referred to as the ``ideal" Higgs Boson scenario. The BABAR~\cite{Lees:2012iw}
and BELLE~\cite{Rorie} 
experiments have placed limits on $m_{a_1}$, using the data collected
at the $\Upsilon$ resonances but it is based on
the ``ideal" Higgs Boson scenario. 
Since in our case the lightest Higgs is around
126 GeV, the  constraints above on  $m_{a_1}$ do not hold. In addition
the LHC experiments~\cite{Chatrchyan:2012am, Chatrchyan:2012cg} have also 
performed a search for a low mass pseudoscalar
$a_1,$ with  $a_1$ decaying to two muons and have obtained the best 
experimental limits till date.

\begin{figure}[htb]
\begin{minipage}[b]{0.45\linewidth}
\centering
\vspace*{0.7cm}
\includegraphics[width=7.5cm, height=5cm]{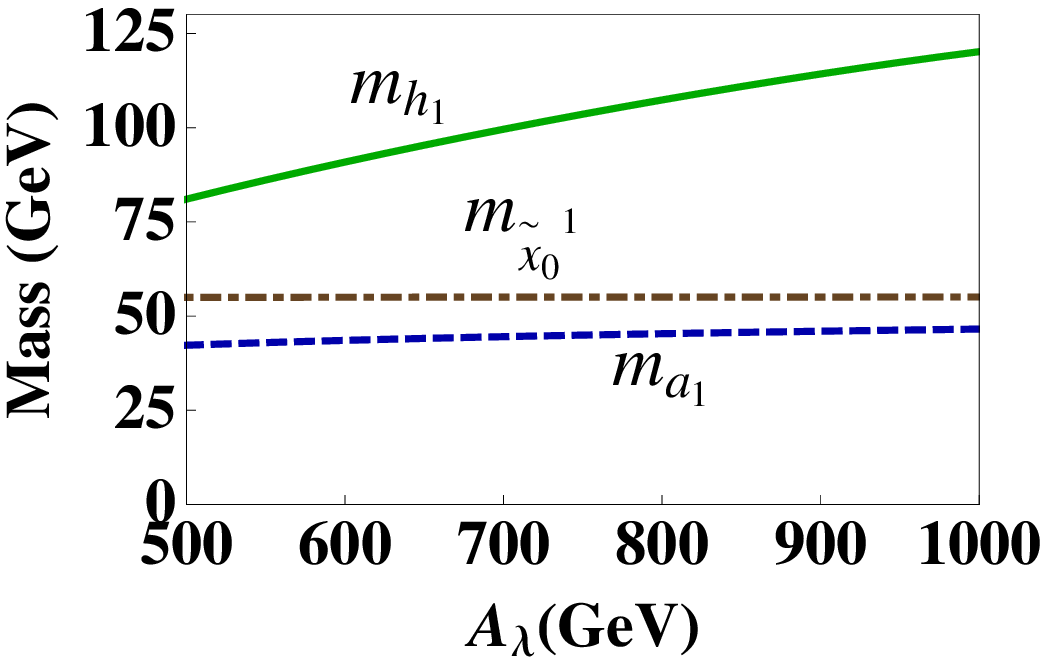}
\caption{Dependence of $m_{h_1}$ [green-solid], $m_{a_1}$ [blue-dashed]
and $m_{\tilde{\chi}_1^0}$ [brown-dot-dashed] on $A_\lambda$
for $M_1$ = 120 GeV, $M_2$ = 200 GeV, $\lambda$ = 0.55 and $\kappa$ = 0.33
with the other input parameters fixed to the values given in 
Table~\ref{tab:nmssmparam}.(colours in on-line version)}
\label{fig:Alambda}
\end{minipage}
\hspace{0.4cm}
\begin{minipage}[b]{0.45\linewidth}
\centering
\includegraphics[width=7.5cm, height=5cm]{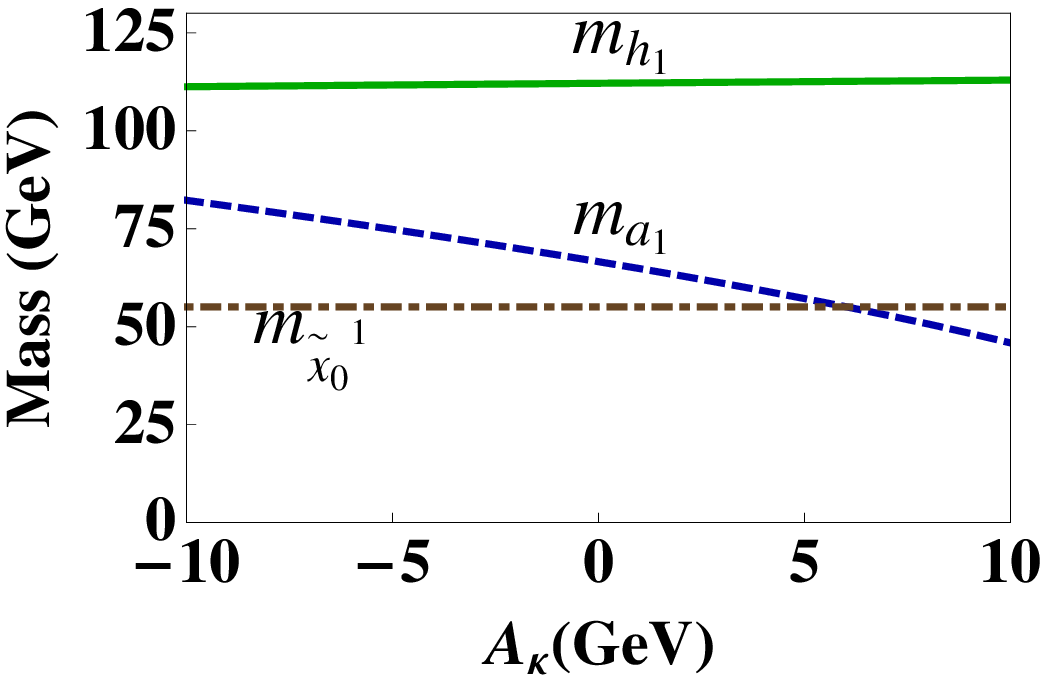}
\caption{Dependence of $m_{h_1}$ [green-solid], $m_{a_1}$ [blue-dashed]
and $m_{\tilde{\chi}_1^0}$ [brown-dot-dashed] on $A_\kappa$
for $M_1$ = 120 GeV, $M_2$ = 200 GeV, $\lambda$ = 0.55 and $\kappa$ = 0.33
with the other input parameters fixed to the values given in 
Table~\ref{tab:nmssmparam}. (colours in on-line version)}
\label{fig:Akappa}
\end{minipage}
\end{figure}

In  Figs.~\ref{fig:mueff},~\ref{fig:tanbeta},~\ref{fig:lambda},
\ref{fig:kappa},~\ref{fig:Alambda}, and ~\ref{fig:Akappa} we show 
the dependence of the mass of $h_1,~a_1,~\tilde{\chi}_1^0$ on 
various parameters of NMSSM. While displaying the dependence on a 
particular parameter, the
other parameters are kept fixed at their  values in Table~\ref{tab:nmssmparam},
with $\lambda$ and $\kappa$ fixed to the lowest acceptable values
of 0.55 and 0.33, respectively. We have fixed the value of the soft gaugino 
mass parameter $M_2 = 200$ GeV, with   $M_1  = 120$ GeV. 
Since the mass $m_{h_1}$ of  CP even,  and 
the mass $m_{a_1}$ of the pseudoscalar  Higgs
are independent of the soft gaugino mass parameters, the dependence
of their mass on various input parameters 
is independent of the  universal gaugino masses at the GUT scale. 
The mass of the lightest neutralino being sensitive to gaugino masses
can be scaled up and down, with its mass as low as 1 GeV for $M_1$
= 5 GeV.  It is seen from 
Figs.~\ref{fig:mueff} and ~\ref{fig:tanbeta} that $m_{h_1}$ and $m_{a_1}$
are sensitive to both $\mu_{eff}$ and $\tan \beta,$ with $m_{a_1}$ being 
comparatively more sensitive. Both these masses  decrease with $\mu_{eff}$.
In case of NMSSM for large $\lambda$, where 
$\lambda \approx$ 0.5 - 0.7, small values of $\tan \beta$
are preferred in order to obtain $m_{h_1}$ in the desired mass window 
of 123 - 127 GeV.
The mass of the lightest neutralino  increases, as expected, with
increasing $\mu_{eff},$ and is almost independent of $\tan \beta$. 
Similarly, we can draw conclusions from Figs.~\ref{fig:lambda},
\ref{fig:kappa},~\ref{fig:Alambda}, and  ~\ref{fig:Akappa}
regarding the dependence of the mass of the 
lightest scalar Higgs, lightest pseudoscalar 
Higgs and the lightest neutralino on different parameters of the 
NMSSM.
Before discussing the branching ratios of the lightest Higgs scalar
to neutralinos and the lightest pseudoscalars, {\it with $M_1$ and $M_2$ 
treated as independent parameters}, in the following  
we summarize the dependence of our results  on the various parameters
of NMSSM:
\begin{itemize}
\item Dependence on $M_1$, $M_2$: 
If the value of $M_1$ is lowered below 30 GeV, the 
neutralino becomes sufficiently light, with 
$h_1 \rightarrow \tilde{\chi}_1^0\tilde{\chi}_1^0$
dominating over the decay $h_1 \rightarrow a_1a_1$
for the entire parameter space considered here. 
If we decrease the  value of
$M_2$, the chargino mass 
bound from the LEP results in larger values of $\mu_{eff}$
being disfavored.
\item Dependence on $\mu_{eff}$: 
Increasing the value of the $\mu_{eff}$, in the
considered range, $m_{a_1}$ reduces whereas $m_{\tilde{\chi}_1^0}$
increases. Therefore the invisible branching ratio 
for $h_1 \rightarrow \tilde{\chi}_1^0\tilde{\chi}_1^0$ decreases, while the 
branching ratio of $h_1 \rightarrow a_1 a_1$ increases.
\item Dependence on $\tan \beta$: 
In this case $h_1 \rightarrow \tilde{\chi}_1^0\tilde{\chi}_1^0$
decreases due to the increase of the branching ratio to lightest pseudoscalars,
as the value of $m_{a_1}$ decreases and $m_{\tilde{\chi}_1^0}$ remains 
constant. 
\item Dependence on $\lambda,~\kappa$: 
Increasing the value of $\lambda$ increases
the value of $m_{a_1}$ and decreases $m_{\tilde{\chi}_1^0}$ .
Since $\lambda$ also substantially affects the mass of $m_{h_1}$,
other parameters need to be changed accordingly, so as to obtain the 
lightest CP even Higgs $h_1$ in the required mass range. The dependence of
$h_1 \rightarrow \tilde{\chi}_1^0\tilde{\chi}_1^0$
on $\lambda$ and $\kappa$ will be discussed in what follows.
\item Dependence on $A_\lambda,~A_\kappa$: 
The pseudoscalar and the neutralino mass is
almost insensitive to $A_\lambda$. We have therefore performed our analyses for
a fixed value of $A_\lambda$ so as to have  $h_1$
in the required mass range. The pseudoscalar mass is  
sensitive to $A_\kappa$, therefore the decay $h_1 \rightarrow a_1 a_1$ can 
be dominant for small $|A_\kappa|$.
\end{itemize}
\begin{figure}[htb]
\includegraphics[width=6.5cm, height=5cm]{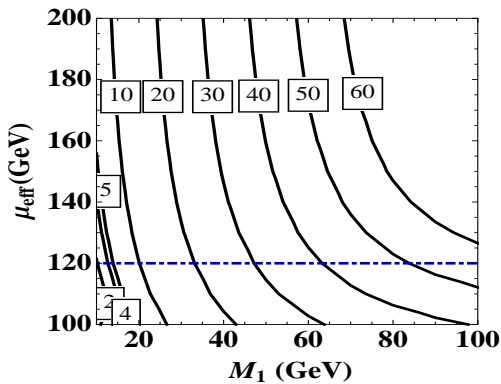}
\caption{Contours of constant lightest neutralino mass $m_{\tilde{\chi}_1^0}$
in the $\mu_{eff} - M_1$ plane for  $M_2$ = 200 GeV,
$\lambda$ = 0.55 and $\kappa$ = 0.6 in NMSSM,
with the other input
parameters fixed at values as given in Table~\ref{tab:nmssmparam}.}
\label{fig:nmssm_neu}
\end{figure}
We now consider the case when the soft
gaugino masses are treated as independent parameters.
In Fig.~\ref{fig:nmssm_neu} we show
contours of constant neutralino mass  in the
$\mu_{eff} - M_1$ plane. We have taken into account the LEP
constraint on the chargino mass ($m_{\tilde{\chi}^\pm} \geq 105$ GeV)
as well as the invisible $Z^0$ decay width (\ref{z_width}). 
For the parameter space considered here, the $Z^0$ invisible decay 
width  is less than 3 MeV. It can be seen from
Fig.~\ref{fig:nmssm_neu} that most of the parameter region with
low $M_1$ allows a low mass neutralino, making the Higgs invisible
decay kinematically possible. The values in this Fig. are obtained  with
$\lambda$ = 0.55 and $\kappa$ = 0.6, with  other parameter values
as given in Table~\ref{tab:nmssmparam}, with $M_2$ = 200 GeV.
The dependence of the constant contours on other parameters can be
inferred from  Figs.~\ref{fig:tanbeta},~\ref{fig:lambda},
\ref{fig:kappa},~\ref{fig:Alambda}, and ~\ref{fig:Akappa}.
\begin{figure}[htb]
\begin{minipage}[b]{0.45\linewidth}
\centering
\includegraphics[width=6.5cm, height=5cm]{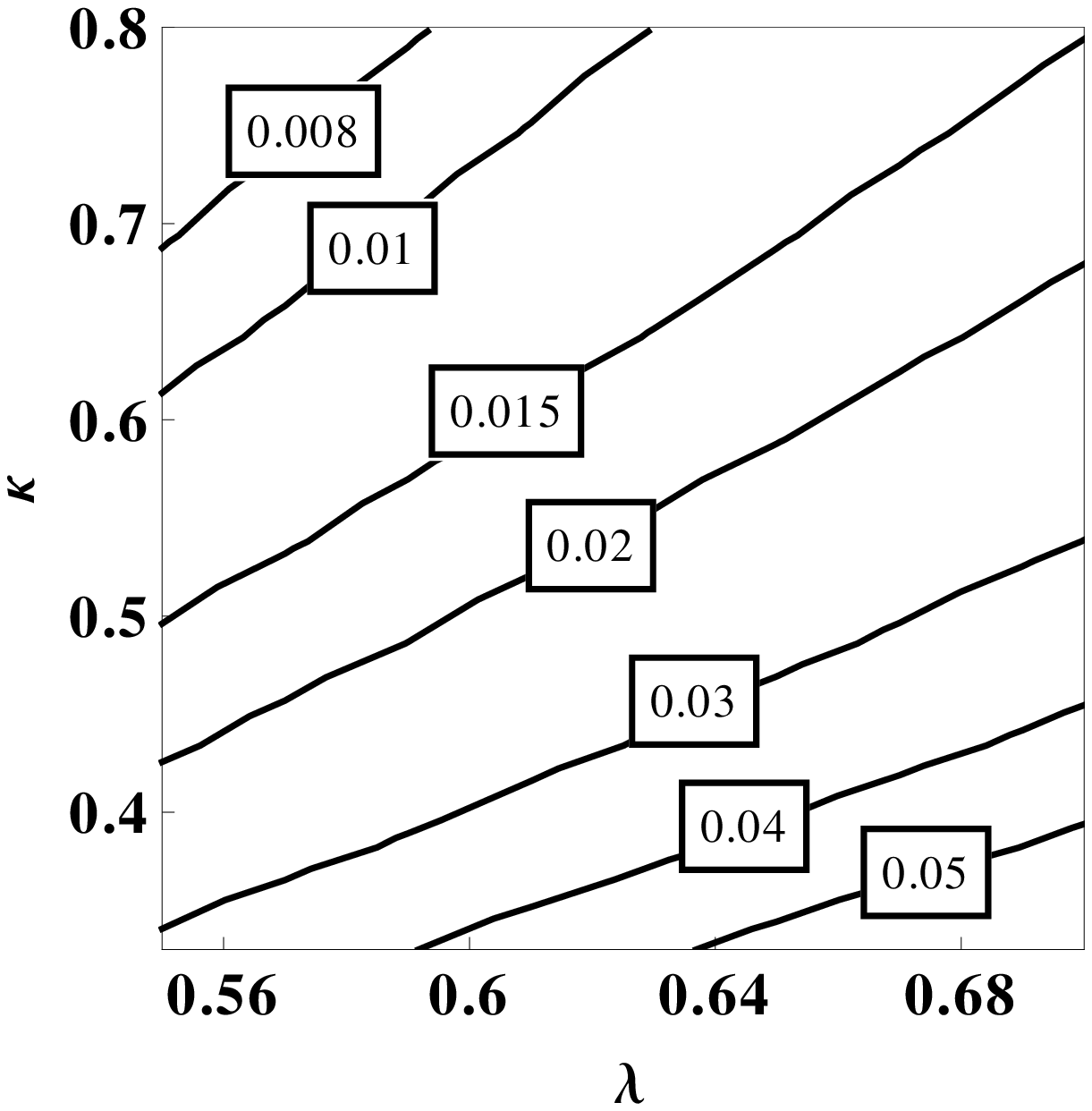}
\caption{Contours of the constant singlino composition 
for NMSSM in the $\kappa - \lambda$
plane for $M_2$ = 200 GeV and $M_1$ = 5 GeV, with the other input
parameters fixed at values as in  Table~\ref{tab:nmssmparam}.}
\label{fig:nmssm_sing1}
\end{minipage}
\hspace{0.4cm}
\begin{minipage}[b]{0.45\linewidth}
\centering
\includegraphics[width=6.5cm, height=5cm]{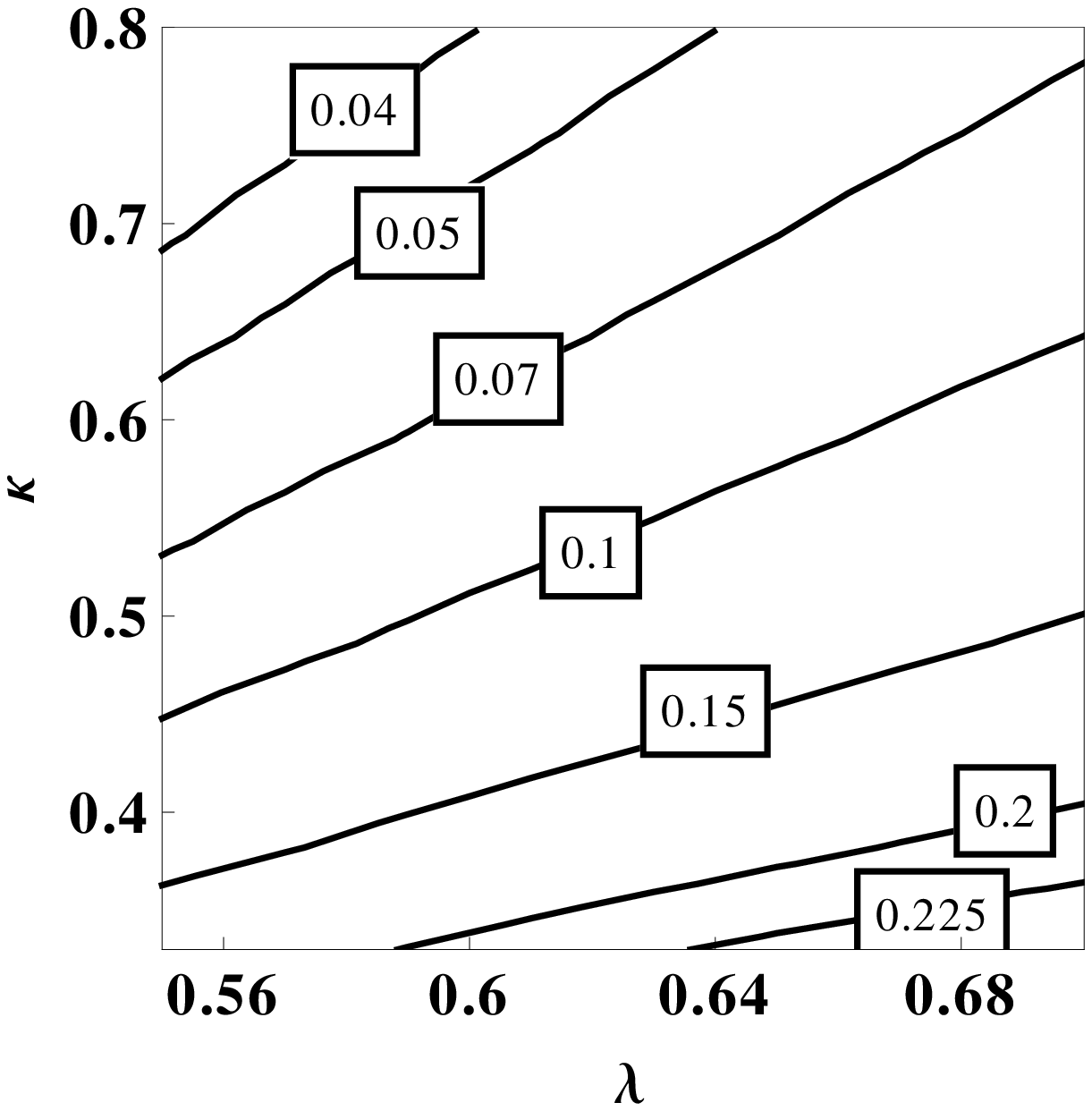}
\caption{Contours of constant singlino composition 
for NMSSM in the $\kappa - \lambda$
plane for $M_2$ = 200 GeV and $M_1$ = 120 GeV, with the other input
parameters fixed at values as in Table~\ref{tab:nmssmparam}.}
\label{fig:nmssm_sing2}
\end{minipage}
\end{figure}  

\begin{figure}[htb]
\begin{minipage}[b]{0.45\linewidth}
\centering
\includegraphics[width=6.5cm, height=5cm]{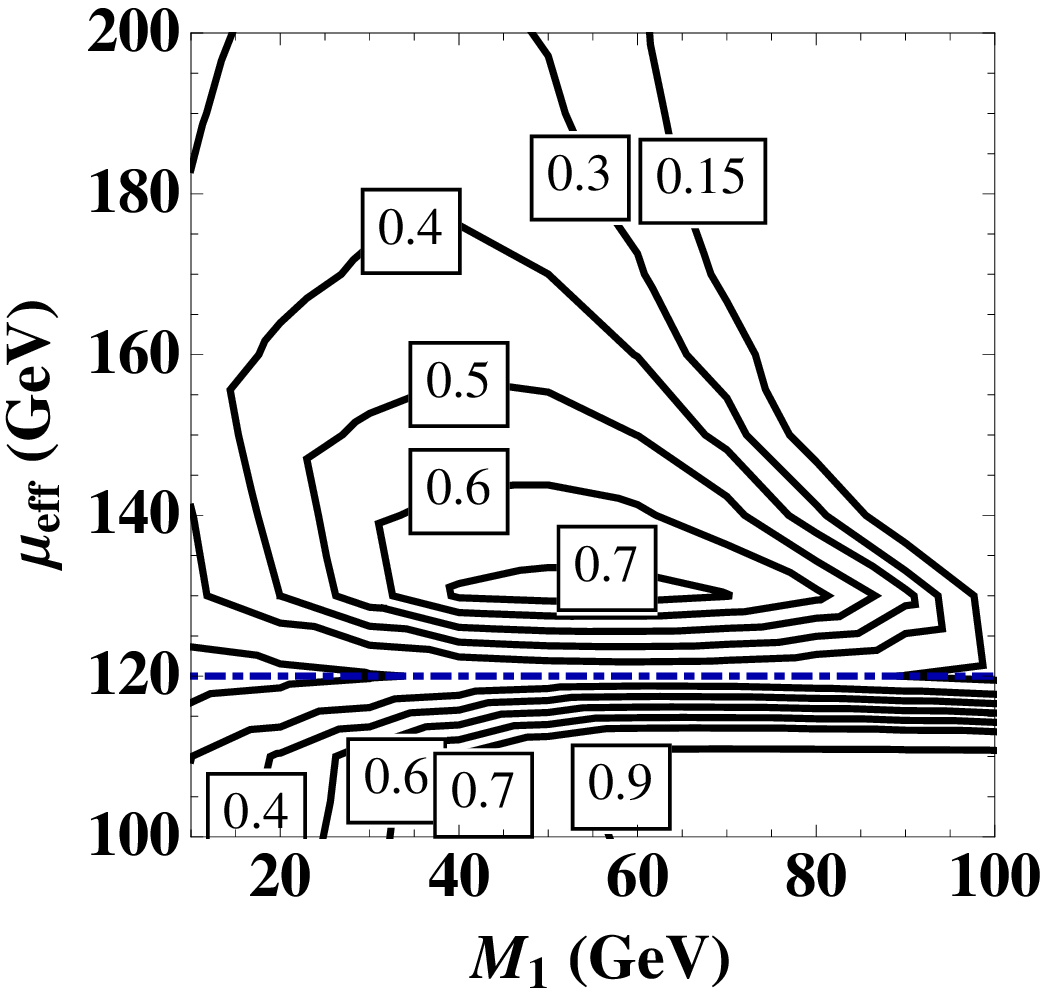}
\caption{Contours of constant branching ratio of 
$(h_1\rightarrow \tilde{\chi}_1^0 \tilde{\chi}_1^0)$
in NMSSM in the  $\mu_{eff} - M_1$ plane
for a fixed value of $M_2$ = 200 GeV, $\lambda$ = 0.55 and $\kappa$ = 0.6,
with the other input
parameters fixed at values as shown in Table~\ref{tab:nmssmparam}.}
\label{fig:nmssm_br}
\end{minipage}
\hspace{0.4cm}
\begin{minipage}[b]{0.45\linewidth}
\centering
\includegraphics[width=6.5cm, height=5cm]{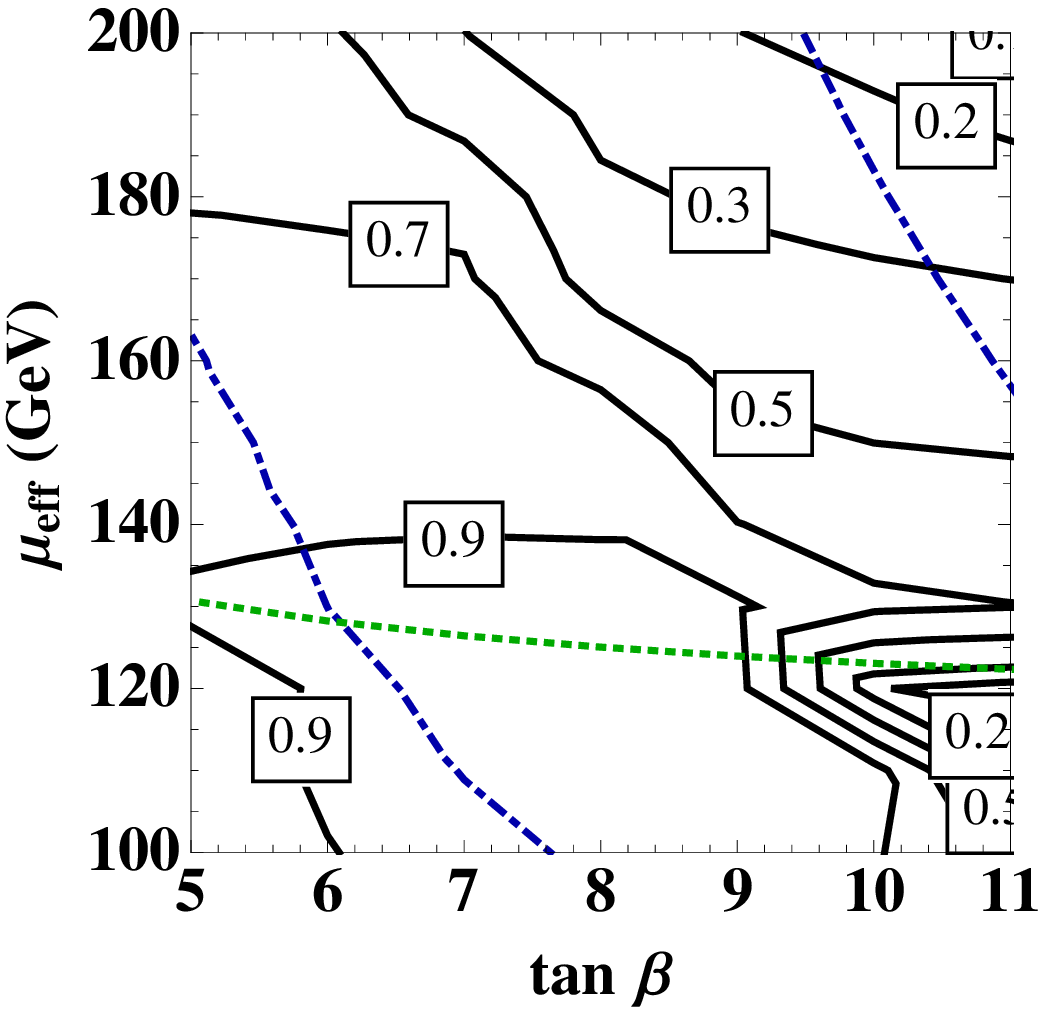}
\caption{Contours of  constant branching ration of 
($h_1\rightarrow \tilde{\chi}_1^0 \tilde{\chi}_1^0$)
in the  $\mu_{eff} - \tan \beta$ plane
for $M_2$ = 200 GeV, $M_1$ = 60 GeV,
$\lambda$ = 0.55 and $\kappa$ = 0.6,
with the other input
parameters fixed at values as given in Table~\ref{tab:nmssmparam}. The
area between the green-dotted lines has $h_1$ in the mass range
123 - 127 GeV.(colours in on-line version)}
\label{fig:mu_tanbeta}
\end{minipage}
\end{figure}

Before considering the  invisible decay width, we  show the contours of 
constant singlino component in
the non GUT scenario, with $M_1$ and $M_2$ treated as independent
parameters.  In Figs.~\ref{fig:nmssm_sing1} and ~\ref{fig:nmssm_sing2} we show
the  constant singlino composition contours
for two different values of $M_1 = 5$ GeV and $120$ GeV,  respectively,
with a fixed value of $M_2 = 200$ GeV. 
The behavior of the constant contours can be 
understood from the fact that for low $M_1$, the neutralino 
is dominantly a gaugino type, with small singlino composition.
Therefore, as discussed earlier, due to the small singlino composition,
the invisible decay width of $h_1$ will be 
large compared to the GUT case. This can be seen in 
Fig.~\ref{fig:nmssm_br},
where we show the invisible branching ratio
of the Higgs decay to the lightest neutralinos in the 
$\mu_{eff} - M_1$ plane. We have fixed $M_2$ = 200 GeV, 
$\lambda$ = 0.55, $\kappa$ = 0.6, with  other input
parameters as  given in Table~\ref{tab:nmssmparam}.
The LEP constraint on the chargino mass excludes the parameter
region below $\mu_{eff}$ = 120 GeV, for $M_2$ = 200 GeV
and is shown by the blue-dot-dashed line. This
limit on $\mu_{eff}$ will decrease, with the increase in the 
value of $M_2$.  The invisible decay width of the $Z^0$ to
the lightest neutralinos satisfies the experimental constraints for the 
entire $\mu_{eff} - M_1$ plane considered here. We see that in the allowed
parameter space, the invisible branching ratio can be as large as 70\%.
The shape of the contours can be understood from Fig.~\ref{fig:mueff},
where we see that $m_{a_1}$ decreases and $m_{\tilde{\chi}_1^0}$
increases, with increasing $\mu_{eff}$, leading to 
$h_1\rightarrow a_1 a_1$ at high $\mu_{eff}$ . At low $\mu_{eff}$ and $M_1$,
$\tilde{\chi}_2^0$ is sufficiently light, therefore the decay 
$h_1\rightarrow \tilde{\chi}_1^0 \tilde{\chi}_2^0$ is kinematically
possible.  This explains the kinks in the contours. 
The  second lightest neutralino $\tilde{\chi}_2^0$ is mostly a higgsino,
at low $\mu_{eff}$ and $M_1$. The bino component increases, 
with the increase in value of $M_1$, for a fixed $\mu_{eff}$.
The dominant branching 
ratio is seen for values of $M_1$ in the range of $40 - 70$ GeV, where 
$m_{\tilde{\chi}_1^0}$ in turn varies from 30 - 60 GeV. In the region
excluded by the chargino mass bound, it is seen that the branching ratio
of Higgs to neutralinos can reach around 90\% for $M_1 >$ 70 GeV and low 
$\mu_{eff}$. This is mainly because in this parameter region both
$m_{\tilde{\chi}_2^0} < (m_{h_1} - m_{\tilde{\chi}_1^0})$ 
and $m_{a_1} <  m_{h_1}/2$. Thus, if the bound on 
invisible branching ratio is considered to be less than 30\%, most
of the region with $\mu_{eff} <$  170 GeV and $M_1 <$ 80 GeV is disfavored
by the invisible Higgs decay. 

In order to fully understand the dependence of the invisible branching ratio
on other input parameters of the NMSSM, in  Fig.~\ref{fig:mu_tanbeta}
we show its behavior in the $\mu_{eff} - \tan \beta$ plane for 
$M_2$ = 200 GeV, $M_1$ = 60 GeV, $\lambda$ = 0.55 and $\kappa$ = 0.6. 
The other input parameters 
are fixed at values in Table~\ref{tab:nmssmparam}. 
We have shown the result for $M_1$ = 60 GeV, as we  see from
Fig.~\ref{fig:nmssm_br}, the dominant branching ratio is seen 
for values of $M_1$ in the region of 40 - 70 GeV. 
The area between the green-dotted lines 
in Fig.~\ref{fig:mu_tanbeta} shows the 
parameter region which allows $h_1$ to be in the allowed
mass range 123 - 127 GeV. The blue-dot-dashed line represents the chargino
mass bound from the LEP. We see that  in the constrained space,
the invisible branching ratio can be as high as 90\%. At small
values of $\tan \beta (< 10)$, when the value of $\mu_{eff}$ is increased,
the invisible branching ratio decreases as $m_{\tilde{\chi}_1^0}$ increases.
The invisible branching ratio is small for $\tan \beta >$ 10 and
low $\mu_{eff}$, due to the opening of the decay channel 
$h_1 \rightarrow a_1a_1$, as $m_{a_1}$ decreases with $\tan\beta$.
This can be seen from Fig.~\ref{fig:tanbeta}. Therefore, considering
the bound on $h_1 \rightarrow \tilde{\chi}_1^0 \tilde{\chi}_1^0$
to be less than 30\%, $\mu_{eff} <$ 180 GeV and $\tan \beta >$ 10 is disfavored.
When $M_1$ is less than 40 GeV, the channel
$h_1\rightarrow \tilde{\chi}_1^0 \tilde{\chi}_2^0$ is kinematically 
accessible for low values of $\mu_{eff}$. The invisible branching 
ratio in this case being small, a large parameter region 
in the $\mu_{eff} - \tan \beta$ plane is favored 
by the bound from LHC experiments. 

The sensitivity of our results on the 
parameters $\lambda$ and $\kappa$ can be understood from the 
behavior of the invisible branching ratio in the $\kappa-\lambda$ plane.
This behavior depends on the composition of the lightest neutralino
and can be easily understood from Figs.~\ref{fig:lambda},~\ref{fig:kappa}.
Since $m_{\tilde{\chi}_1^0}$ is sensitive to the gaugino mass parameter
$M_1$, we discuss the behavior for different values of $M_1$. At low 
values of $M_1 <$ 30 GeV, as discussed earlier, the channel 
$h_1\rightarrow \tilde{\chi}_1^0\tilde{\chi}_2^0$ becomes kinematically
accessible. Therefore the Higgs invisible branching ratio is less
than 30\% for most of the $\kappa-\lambda$ parameter space.
With 40 $< M_1 <$ 70 GeV, as can be seen from Fig.~\ref{fig:nmssm_br},
the branching ratio of $h_1\rightarrow \tilde{\chi}_1^0 \tilde{\chi}_1^0$
is the largest.  As $m_{\tilde{\chi}_1^0}$ decreases with 
$\lambda$, Fig.~\ref{fig:lambda}, the neutralino becomes light 
($m_{\tilde{\chi}_1^0} < m_{h_1}/2$), and the mass of the lightest
pseudoscalar Higgs increases ($m_{a_1} >  m_{h_1}/2$), with 
$\lambda >$ 0.6, even in case of large $M_1$.
Therefore the dominant decay
mode is $h_1\rightarrow \tilde{\chi}_1^0 \tilde{\chi}_1^0$, with the branching 
ratio greater the 90\% for $\lambda >$ 0.6. This result is practically
independent of $\kappa$ as can be seen from Fig.~\ref{fig:kappa},
where $m_{\tilde{\chi}_1^0}, m_{a_1}$ is seen not to  depend on $\kappa$.
Again for large $M_1$ and $\lambda <$ 0.6, with $m_{a_1}<m_{\tilde{\chi}_1^0}$, 
the branching ratio of $h_1 \rightarrow \tilde{\chi}_1^0 \tilde{\chi}_1^0$
is smaller. 

The invisible decay width mainly depends on the neutralino composition. The
neutralino should have a small singlino component and a dominant bino
component, i.e. $M_1$ should be small, in order to have a large invisible
decay width. The 
decay width is also sensitive to the mass of the pseudoscalar Higgs $a_1$ and 
next to the lightest neutralino $\tilde{\chi}_2^0$. The dependence
of the width on the 
other input parameters  $\tan \beta$, $\kappa$ and $\lambda$
is sensitive to the gaugino mass parameter $M_1$ and behaves differently
for smaller  and larger  values of $M_1$. This is mainly because 
$m_{\tilde{\chi}_2^0}$ is also sensitive to $M_1$, leading to the opening 
of new decay channels. 

 \section{Summary and Conclusions}
 \label{sec:conclusions}
 
We now summarize the results obtained in this work.  
We have considered the possibility of the invisible
decays of the lightest CP
even Higgs boson in  MSSM and in  NMSSM.  In the MSSM, we have considered
both the universal as well as nonuniversal gaugino masses at the GUT scale.
In both cases we have seen that it is not possible to have a light neutralino,
so that the decay of the lightest Higgs boson to lightest neutralinos
does not take place. Our results show that in virtually all realistic
scenarios, the non-universality is not sufficient to generate
sufficiently light neutralinos.  We have parametrized such non-universality
in terms of a parameter $r$ which we have studied in detail.  The details
of such non-universality are briefly summarized in the 
Appendix~\ref{appendix: nonuniversal}.

We have then analyzed the possibility of having a light neutralino
in the NMSSM extension of MSSM.
We note that in the NMSSM, both the Higgs as well as the neutralino
sectors are significantly richer, which  provides us with
greater possibilities. We have considered the neutralino sector of NMSSM,
and in particular the phenomenon
of the mixing of the singlino, and  concluded that even in this case
massless neutralinos cannot be realized with
universal boundary conditions on the gaugino masses at the GUT scale,
since the lightest Higgs is
too heavy in conflict with the LHC result.  
Furthermore, with universal boundary conditions the lightest Higgs
$m_{h_1} \approx 126$ GeV would not decay to the lightest neutralinos. 
%
A related consideration is the
``ideal" Higgs scenario motivated by LEP constraints, where the 
next to lightest CP even Higgs $h_2$ can decay to lightest neutralinos.
Departing from the assumption of universal gaugino masses, we have
investigated  the invisible branching ratio of the Higgs, as a function
of the various  parameters of NMSSM. We have concentrated on the 
case with the lightest scalar as the SM Higgs boson $h_1$, and have 
considered the dependence on parameters which are 
relevant to the Higgs and neutralino sector.

As is well known, the Higgs sector of the NMSSM itself is richer
than the corresponding one in the MSSM.  Thus, there is the intriguing
possibility that the Higgs can  decay into a pair of CP odd
lightest Higgs particles $a_1$.  It is seen that for higher values of
$\tan \beta$, the invisible branching ratio decreases, with the largest
contribution coming from the Higgs decaying to two light pseudoscalar 
Higgs bosons.
The present Higgs decay uncertainties can
constrain NMSSM but these constraints are strongly correlated with the 
composition of the lightest neutralino. The invisible branching ratio
is found to be relatively independent of $\lambda$ and $\kappa$, for
40 GeV $< M_1 <$ 60 GeV. In the NMSSM, the constraints on 
the Higgs mass results in small values of $\tan \beta$ 
being favored for large $\lambda$. 
We have discussed the dependence of our results on the parameters
which enter the neutralino and the Higgs sectors of the NMSSM.
From the dependence of the invisible branching ratio 
in the $\mu_{eff}-M_1$ plane, with other parameters  fixed, 
we have shown that most of the parameter space is constrained by 
considering invisible 
branching ratio $<$ 30\%.
The dependence of this result on the other input parameters has 
also been discussed. For large values of  $\tan \beta$, 
the invisible branching ratio
decreases as $a_1$ becomes lighter with $h_1\rightarrow a_1a_1$ kinematically
possible. Therefore at large $\tan \beta$, $M_1 <$ 40 GeV is favored in the 
$\mu_{eff}-M_1$ plane, for all values of $\mu_{eff}$. The allowed parameter
region with $M_1>$  80 GeV remains unchanged. The sensitivity of the results 
on the input parameters $\lambda,~\kappa$ has also been discussed in detail.
We have shown that 
for $M_1 <$ 70 GeV, the results do not change significantly as a function of
of $\lambda$ and $\kappa$.
But with large $M_1$ and $\lambda >$ 0.6, the neutralinos becomes very light.
In that case the $\mu_{eff}-M_1$ parameter space is more tightly constrained.
Further  data from LHC may be able to shed light on the question of the
invisible decays of the lightest Higgs boson.
%
\section{acknowledgements} PNP would like to thank the
Centre for High Energy Physics, Indian Institute 
of Science, Bangalore for  hospitality while this work was initiated, 
as well as the Inter-University Centre for Astronomy and 
Astrophysics, Pune, for hospitality where part of this work was done. The 
work of PNP is supported by the J. C. Bose National Fellowship 
of the Department of 
Science and Technology, India, and by the Council of Scientific and 
Industrial Research, India.

\appendix
\section{Nonuniversal gaugino masses in GUTS}
\label{appendix: nonuniversal}
In this Appendix we briefly discuss non universal gaugino masses as they
arise in grand unified models~\cite{Pandita:2012es}.
In grand unified supersymmetric models, non-universal gaugino masses 
are generated by a non-singlet chiral superfield $\Phi^n$ that appears 
linearly in the gauge kinetic function 
$f(\Phi)$, which is an analytic function of the 
chiral superfields $\Phi$ in the theory \cite{Cremmer:1982wb}.
The gaugino masses are generated from the coupling of 
the field strength superfield $W^a$ with $f(\Phi)$,
when the auxiliary part $F_\Phi$ of a chiral superfield $\Phi$  in 
$f(\Phi)$ gets a VEV.
The Lagrangian for the coupling of gauge kinetic function to the 
gauge field strength can be  written as
\bea 
{\cal L}_{g.k.} \; & = & \;
\int d^2\theta f_{ab}(\Phi) W^{a}W^{b}
+h.c.,
\label{gk}
\eea
where $a$ and $b$ refer to gauge group indices, and repeated indices are summed 
over. The gauge kinetic function $f_{ab}(\Phi)$ is given by
\bea
f_{ab}(\Phi) & = & f_0(\Phi^s)\delta_{ab} 
+ \sum_n f_n(\Phi^s){\Phi_{ab}^n\over M_P} + 
\cdot \cdot \cdot \cdot \cdot \cdot. 
\eea
Here  $\Phi^s$ and the  $\Phi^n$ are the singlet and 
the non-singlet chiral superfields, respectively. Furthermore,
$f_0(\Phi^s)$ and $f_n(\Phi^s)$ are functions of gauge singlet
superfields $\Phi^s$, and $M_P$ denotes  some large scale.  When $F_\Phi$ 
gets a VEV $\langle F_\Phi \rangle$, the interaction~(\ref{gk}) generates
gaugino masses:  
\bea
{\cal L}_{g.k.} \; \supset \;
{{{\langle F_\Phi \rangle}_{ab}} \over {M_P}}
\lambda^a \lambda^b +h.c., 
\eea
where $\lambda^{a,b}$ are gaugino fields. Here, we  denote 
by $\lambda^1$, $\lambda^2$ and $\lambda^3$  as the 
$U(1)$, $SU(2)$ and $SU(3)$ 
gaugino fields, respectively. Since the gauginos belong to the adjoint 
representation of the gauge group, 
$\Phi$ and $F_\Phi$ can belong to any of the 
representations appearing in the symmetric product of the 
two adjoint  representations of unified gauge group. 

\medskip

In the case where the  SM gauge group 
is embedded with in the  grand unified gauge group $SU(5)$.
For the symmetric product of the two adjoint~({\bf 24} dimensional)
representations of $SU(5)$, we have
\bea
({\bf 24 \otimes 24})_{Symm} = {\bf 1 \oplus 24 \oplus 75 \oplus 200}.
\label{product}
\eea

In Table~\ref{tab1} we show the ratios of  gaugino masses 
which result when $F_{\Phi}$ belongs to different representations of 
$SU(5)$ in the decomposition~(\ref{product}).


\begin{table}[t!]
\renewcommand{\arraystretch}{1.0}
\begin{center}
  \begin{tabular}{||c|ccc|ccc||}
    \hline 
    $SU(5)$ & $M_1^G$ & $M_2^G$ & $M_3^G$ & 
    $M_1^{EW}$ & $M_2^{EW}$ & $M_3^{EW}$
    \\ \hline 
    {\bf 1} & 1 & 1
    & 1 & 1 & 2 & 7.1 \\ 
    & & & & & & \\    
    {\bf 24} & 1 & 3 & -2 & 1 & 6 & -14.3 \\
     & & & & & & \\    
     {\bf 75} & 1 &-$\frac{3}{5}$ &-$\frac{1}{5}$ & 1 & -1.18 & -1.41 \\
      & & & & & & \\    
      {\bf 200} & 1 & $\frac{1}{5}$ &$\frac{1}{10}$ &1 & 0.4 & 0.71
    \\ \hline
  \end{tabular}
  \end{center}
  \caption{\label{tab1} Ratios of the gaugino masses at the GUT scale
    in the normalization ${M_1}(GUT)$ = 1, and at the electroweak
    scale in the normalization ${M_1}(EW)$ = 1 
    for $F$-terms in different representations of $SU(5)$.
    These results are obtained by using 1-loop renormalization
    group equations.}
\renewcommand{\arraystretch}{1.0}
\end{table}
\noindent

\medskip
Next we consider the embedding of the SM gauge group in a $SO(10)$
grand unified theory. The adjoint representation
of $SO(10)$ being ({\bf 45}), $\Phi$ and $F_\Phi$ can belong to
the symmetric product of two 
adjoint ({\bf 45}) dimensional representations~\cite{Martin:2009ad}

\bea 
({\bf 45} \times {\bf 45})_{Symm}={\bf 1} 
\oplus {\bf 54} \oplus {\bf 210} \oplus {\bf 770}.
\label{symmetric_SO10}
\eea


\begin{table}[ht]
  \begin{tabular}{||c|c|ccc|ccc||}
   \hline 
   $SO(10)$ & $SU(5)$ & $M_1^G$ & $M_2^G$ & $M_3^G$ & 
    $M_1^{EW}$ & $M_2^{EW}$ & $M_3^{EW}$
    \\ \hline 
    {\bf 1} & {\bf 1} &1 &1 &1 &1 &2 &7.1\\[0.5 mm]
     {\bf 54} & {\bf 24} &1 &3 &-2 
     & 1 &6  &-14.3 \\[0.5 mm]
 {\bf 210} & {\bf 1}  &1 &1 &1 &1 &2 &7.1\\[0.5 mm]
           &  {\bf 24} &1 &3 &-2
     & 1 & 6  &-14.3 \\[0.5 mm]
           & {\bf 75} & 1 &-$\frac{3}{5}$ &-$\frac{1}{5}$ & 1 & -1.18 & -1.41 \\[0.5 mm]
  {\bf 770} & {\bf 1}  &1 &1 &1 &1 &2 &7.1\\[0.5 mm]
             &  {\bf 24} &1 &3 &-2
     & 1 &6  &-14.3 \\[0.5 mm]
           & {\bf 75} & 1 &-$\frac{3}{5}$ &-$\frac{1}{5}$ & 1 & -1.18 & -1.14 \\[0.5 mm]
           & {\bf 200} & 1 & $\frac{1}{5}$ &$\frac{1}{10}$ &1 & 0.4 & 0.71          
         \\ \hline
  \end{tabular}
   \caption{\label{tab2}Ratios of the gaugino masses at the GUT scale
    in the normalization ${M_1}(GUT)$ = 1, and at the electroweak
    scale in the normalization ${M_1}(EW)$ = 1 
    for $F$-terms in representations of $SU(5)\subset SO(10)$
    with the normal (nonflipped) embedding. These results have been 
    obtained at the 1-loop level.}

\begin{center}
  \begin{tabular}{||c|c|ccc|ccc||}
   \hline 
   $SO(10)$ & $[SU(5)' \times U(1)]_{flipped}$ & $M_1^G$ & $M_2^G$ & $M_3^G$ & 
    $M_1^{EW}$ & $M_2^{EW}$ & $M_3^{EW}$
    \\ \hline 
    {\bf 1} & ({\bf 1},0) &1 &1 &1 &1 &2 &7.1\\ [0.5 mm]
     {\bf 54} & ({\bf 24},0) & 1 &3 &-2 
     & 1 &6  &-14.3 \\ [0.5 mm]
    {\bf 210} & ({\bf 1},0)  & 1 &-$\frac{5}{19}$ &-$\frac{5}{19}$ &1 &-0.52 &-1.85\\ [0.5 mm]
           &  ({\bf 24},0) & 1 &-$\frac{15}{7}$ &$\frac{10}{7}$ 
     & 1 &-4.2  &10 \\ [0.5 mm]
           & ({\bf 75},0) & 1 &-15 & -5 & 1 &-28 & -33.33 \\ [0.5 mm]
  {\bf 770} & ({\bf 1},0)  & 1 &$\frac{5}{77}$  &$\frac{5}{77}$ &1 &0.13 &0.46\\ [0.5 mm]
             &  ({\bf 24},0) & 1 &$\frac{15}{101}$ & -$\frac{10}{101}$  
     & 1 &0.3  &-0.70 \\ [0.5 mm]
           & ({\bf 75},0) & 1 &-15 &-5 & 1 &-28 &-33.3 \\ [0.5 mm]
           & ({\bf 200},0) & 1 &5 & $\frac{5}{2}$ &1 & 9.33 & 16.67          
         \\ \hline
  \end{tabular}
  \end{center}
  \caption{\label{tab3}Ratios of the gaugino masses at the GUT scale
    in the normalization ${M_1}(GUT)$ = 1, and at the electroweak
    scale in the normalization ${M_1}(EW)$ = 1 at the 1-loop level
    for $F$-terms in representations of flipped
    $SU(5)'\times U(1)$ $\subset SO(10)$.}

\vspace{1.0cm}

\begin{minipage}[b]{0.55\linewidth}
\begin{center}
   \begin{tabular}{||c|c|ccc|ccc||}
   \hline 
   $SO(10)$ & $ SU(4) \times SU(2)_R $ & 
                                          $M_1^G$ & $M_2^G$ & $M_3^G$ & 
    $M_1^{EW}$ & $M_2^{EW}$ & $M_3^{EW}$
    \\ \hline 
    {\bf 1} & ({\bf 1},{\bf 1}) &1 &1 &1 &1 &2 &7.1\\ [0.5 mm]
     {\bf 54} & ({\bf 1},{\bf 1}) & 1 &3 &2  
     & 1 &6  &-14.3 \\ [0.5 mm]
    {\bf 210} & ({\bf 1},{\bf 1})  & 1 &-$\frac{5}{3}$ &0 &1 &-3.35 &0\\ [0.5 mm]
           &  ({\bf 15},{\bf 1}) & 1 &0 &-$\frac{5}{4}$
     & 1 &0  &-9.09 \\ [0.5 mm]
           & ({\bf 15},{\bf 3}) & 1 & 0 & 0 &1  & 0 &0 \\ [0.5 mm]
  {\bf 770} & ({\bf 1},{\bf 1})  & 1 &$\frac{25}{19}$ &$\frac{10}{19}$&1 &2.6 &3.7\\ [0.5 mm]
             &  ({\bf 1},{\bf 5}) & 1 &0 &0  
     & 1 &0  &0 \\ [0.5 mm]
           & ({\bf 15},{\bf 3}) & 1 & 0 & 0 & 1 & 0 & 0 \\ [0.5 mm]
           & ({\bf 84},{\bf 1}) & 1 & 0 & $\frac{5}{32}$ &1 & 0 & 1.11          
         \\ \hline 
  \end{tabular}
  \end{center}
   \caption{\label{tab4}Ratios of the gaugino masses at the GUT scale
    in the normalization ${M_1}(GUT)$ = 1, and at the electroweak
    scale in the normalization ${M_1}(EW)$ = 1 at the 1-loop level
    for $F$-terms in representations of 
    $SU(4)\times SU(2)_L \times SU(2)_R \subset SO(10)$.} 
  \end{minipage}
  \end{table}
%
 
\noindent In Table~\ref{tab2} we have shown the gaugino mass
parameters for the different representations that arise in the symmetric
product~(\ref{symmetric_SO10}) for the $SO(10)$ group. We note
from Table~\ref{tab2} that the ratios of gaugino masses
for the different representations of $SO(10)$ in the symmetric 
product (\ref{symmetric_SO10}) with the unflipped embedding 
$SU(5) \subset SO(10)$ are identical to the corresponding gaugino
mass ratios in Table~\ref{tab1} for the embedding of SM in $SU(5)$.
In case of the flipped embedding $SU(5)'\times U(1) \subset SO(10)$,
as seen from Table~\ref{tab3}, the gaugino mass ratios for the 
${\bf 210}$ and ${\bf 770}$ dimensional representations of the 
grand unified gauge groups is different from the corresponding 
ratios for $SU(5)$. The ratio $r$, used for our analyses in
Section~\ref{subsec:nonuniversal gaugino masses} is obtained in this case from
the Tables~\ref{tab2},~\ref{tab3},~\ref{tab4} respectively.


\medskip

Finally we consider the grand unified group $E_6$, which has 
$\bf{78}$ as the adjoint representation~\cite{Martin:2009ad}.
The possible $E_6$ symmetric irreducible representations are
\begin{equation}
 ({\bf 78} \times {\bf 78})_{Symm} = {\bf 1} \oplus {\bf 650} \oplus {\bf 2430}.
 \label{symmetric_E6}
\end{equation}
The corresponding quantities  of interest for this case
are tabulated in Tables~\ref{tab5}, \ref{tab6}, \ref{tab7},
\ref{tab8}, \ref{tab9} and \ref{tab10}.

\begin{table}[htb]
\begin{center}
   \begin{tabular}{||c|c|c|ccc|ccc||}
   \hline 
   $E_6$ & $ [SO(10)' \times U(1)]_{flipped} $ &$SU(5)''$ 
                                          &$M_1^G$ & $M_2^G$ & $M_3^G$ & 
    $M_1^{EW}$ & $M_2^{EW}$ & $M_3^{EW}$
    \\ \hline 
    {\bf 1} & ({\bf 1},0) &{\bf 1} &1 &1 &1 &1 &2 &7.1\\ [0.5 mm]
     {\bf 650} & ({\bf 1},0) & {\bf 1} &1 &-$\frac{5}{22}$ & -$\frac{5}{22}$
     & 1 &-0.46  &-1.61 \\ [0.5 mm]
     & ({\bf 45},0) & {\bf 1} &1 &0 & 0 & 1 &0  &0 \\ [0.5 mm]  
      &             & {\bf24} &1 &0 & 0 & 1 &0  &0 \\ [0.5 mm]  
      & ({\bf 54},0) & {\bf 24} &1 &-15 & 10 & 1 &-30  &70 \\ [0.5 mm]  
      & ({\bf 210},0) & {\bf 1} &1 &-5 & -5 & 1 &-10.0  &-35.5 \\ [0.5 mm]
      &      & {\bf24} &1 &$\frac{15}{2}$ & -5 & 1 &15.1  &-35.5 \\ [0.5 mm]  
       &      & {\bf75} &1 &-15 & -5 & 1 &-30.1  &-35.5 \\ [0.5 mm]
  {\bf 2430} & ({\bf 1},0) & {\bf 1} &1 &$\frac{5}{122}$ & $\frac{5}{122}$
     & 1 &0.08  &0.29 \\ [0.5 mm]
     & ({\bf 45},0) & {\bf 1} &1 &0 & 0 & 1 &0  &0 \\ [0.5 mm] 
      &             & {\bf24} &1 &0 & 0 & 1 &0  &0 \\ [0.5 mm]
      & ({\bf 210},0) & {\bf 1} &1 &-5 & -5 & 1 &-10.0  &-35.5 \\ [0.5 mm]
      &      & {\bf24} &1 &$\frac{15}{2}$ & -5 & 1 &15.1  &-35.5 \\ [0.5 mm]  
      &      & {\bf75} &1 &-15 & -5 & 1 &-30.1  &-35.5 \\ [0.5 mm]
      & ({\bf 770},0) & {\bf 1} &1 &1 & 1 & 1 &2  &7.1 \\ [0.5 mm]
      &   & {\bf24} &1 &-$\frac{3}{5}$ & $\frac{2}{5}$ & 1 &-1.21 &2.84 \\ [0.5 mm] 
      &      & {\bf75} &1 &-15 & -5 & 1 &-30.1  &-35.5 \\ [0.5 mm]
      &    &{\bf 200} &1 &5 &$\frac{5}{2}$ &1 &10.0 &17.7 \\ \hline 
  \end{tabular}
  \end{center}
   \caption{\label{tab5}Ratios of the gaugino masses at the GUT scale
    in the normalization ${M_1}(GUT)$ = 1, and at the electroweak
    scale in the normalization ${M_1}(EW)$ = 1 at the 1-loop level
    for $F$-terms in representations of 
    $SU(5)''\times U(1)' \times U(1) \subset [SO(10)' \times U(1)]_{flipped}
    \subset E_6$.} 
 
  \end{table}

\begin{table}[htb]
\begin{center}
   \begin{tabular}{||c|c|c|ccc|ccc||}
   \hline 
   $E_6$ & $ [SO(10)' \times U(1)]_{flipped} $ &$SU(4)'$ 
                                          &$M_1^G$ & $M_2^G$ & $M_3^G$ & 
    $M_1^{EW}$ & $M_2^{EW}$ & $M_3^{EW}$
    \\ \hline 
    {\bf 1} & ({\bf 1},0) &{\bf 1} &1 &1 &1 &1 &2 &7.1\\ [0.5 mm]
     {\bf 650} & ({\bf 1},0) & {\bf 1} &1 &-$\frac{5}{22}$ & -$\frac{5}{22}$
     & 1 &-0.46  &-1.61 \\ [0.5 mm]
     & ({\bf 45},0) & {\bf 15} &1 &0 & 0 & 1 &0  &0 \\ [0.5 mm]  
     & ({\bf 54},0) & {\bf 1} &1 &-15 & 10 & 1 &-30  &70 \\ [0.5 mm]  
      & ({\bf 210},0) & {\bf 1} &0 &1 & 0 & 0 &1  &0 \\ [0.5 mm]
      &      & {\bf15} &1 &0 & -5 & 1 &0  &-35.5 \\ [0.5 mm]  
  {\bf 2430} & ({\bf 1},0) & {\bf 1} &1 &$\frac{5}{122}$ & $\frac{5}{122}$
     & 1 &0.08  &0.29 \\ [0.5 mm]
     & ({\bf 45},0) & {\bf 15} &1 &0 & 0 & 1 &0  &0 \\ [0.5 mm] 
     & ({\bf 210},0) & {\bf 1} &0 &1 & 0 &0 &1  &0 \\ [0.5 mm]
     &      & {\bf15} &1 &0 & -5 & 1 &0  &-35.5 \\ [0.5 mm]  
     & ({\bf 770},0) & {\bf 1} &1 &25 & 10 & 1 &50.2  &70.9 \\ [0.5 mm]
     &   & {\bf84} &1 &0 & $\frac{5}{8}$ & 1 &0 &4.43 \\ \hline 
  \end{tabular}
  \end{center}
   \caption{\label{tab6}Ratios of the gaugino masses at the GUT scale
    in the normalization ${M_1}(GUT)$ = 1, and at the electroweak
    scale in the normalization ${M_1}(EW)$ = 1 at the 1-loop level
    for $F$-terms in representations of 
    $SU(4)'\times SU(2)_L \times SU(2)_X \times U(1) 
    \subset [SO(10)' \times U(1)]_{flipped}
   \subset E_6$ } 
  \end{table}

\begin{table}[htb]
\begin{center}
   \begin{tabular}{||c|c|ccc|ccc||}
   \hline 
   $E_6$ & $ SU(3)_L \times SU(3)_R $ & 
                                          $M_1^G$ & $M_2^G$ & $M_3^G$ & 
    $M_1^{EW}$ & $M_2^{EW}$ & $M_3^{EW}$
    \\ \hline 
    {\bf 1} & ({\bf 1}, {\bf 1}) &1 &1 &1 &1 &2 &7.1\\ [0.5 mm]
     {\bf 650} & $({\bf 1},{\bf 1})_1$ & 1 &-$\frac{5}{3}$ &0  
     & 1 &-3.35  &0 \\ [0.5 mm]
     & $({\bf 1},{\bf 1})_2$ & 1 &0 & -$\frac{5}{4}$ & 1 &0  &-8.86 \\ [0.5 mm]
     & ({\bf 1}, {\bf 8}) & 1 &0 &0 & 1 &0  &0 \\ [0.5 mm]
      & ({\bf 8}, {\bf 1}) & 1 &-5 &0 & 1 &-10.0  &0 \\ [0.5 mm]
      & ({\bf 8}, {\bf 8}) & 1 &0 &0 & 1 &0  &0 \\ [0.5 mm] 
    {\bf 2430} & ({\bf 1}, {\bf 1})  & 1 &1 &1 &1 &2 &7.1 \\ [0.5 mm]
    & ({\bf 1}, {\bf 8}) & 1 &0 &0 & 1 &0  &0 \\ [0.5 mm]
      & ({\bf 8}, {\bf 1}) & 1 &-5 &0 & 1 &-10.0  &0 \\ [0.5 mm]
      & ({\bf 8}, {\bf 8}) & 1 &0 &0 & 1 &0  &0 \\ [0.5 mm] 
     & ({\bf 1}, {\bf 27}) & 1 &0 &0 & 1 &0  &0 \\ [0.5 mm]
      & ({\bf 27}, {\bf 1}) & 1 &$\frac{5}{9}$ &0 & 1 &1.12  &0  \\ \hline 
  \end{tabular}
  \end{center}
   \caption{\label{tab7}Ratios of the gaugino masses at the GUT scale
    in the normalization ${M_1}(GUT)$ = 1, and at the electroweak
    scale in the normalization ${M_1}(EW)$ = 1 at the 1-loop level
    for $F$-terms in representations of the trinification subgroup
    $SU(3)_C \times SU(3)_L \times SU(3)_R \subset E_6$.} 
   \end{table}
%

\begin{table}[htb]
\begin{center}
   \begin{tabular}{||c|c|c|ccc|ccc||}
   \hline 
   $E_6$ & $ SU(6) \times SU(2)_X $ &$SU(3)_L$ & 
                                          $M_1^G$ & $M_2^G$ & $M_3^G$ & 
    $M_1^{EW}$ & $M_2^{EW}$ & $M_3^{EW}$
    \\ \hline 
    {\bf 1} & ({\bf 1}, {\bf 1}) &{\bf 1} &1 &1 &1 &1 &2 &7.1\\ [0.5 mm]
     {\bf 650} & ({\bf 1}, {\bf 1})&{\bf 1} & 1 &1 &1  
     & 1 &2 &7.1 \\ [0.5 mm]
     & ({\bf 35}, {\bf 1}) &{\bf 1} & 1 &5 & -5 & 1 &10.0  &-35.5 \\ [0.5 mm]
     &  &{\bf 8} & 1 &$\frac{5}{3}$ & 0 & 1 &3.35  &0 \\ [0.5 mm]
     & ({\bf 189}, {\bf 1}) &{\bf 1} & 1 &-$\frac{1}{3}$ &-$\frac{1}{3}$
     & 1 &-0.67  &-2.36 \\ [0.5 mm]
     &  &{\bf 8} & 1 &-1 & 0 & 1 &-2  &0 \\ [0.5 mm]
  {\bf 2430} & ({\bf 1}, {\bf 1})&{\bf 1} & 1 &1 &1  
     & 1 &2 &7.1 \\ [0.5 mm]    
   & ({\bf 189}, {\bf 1}) &{\bf 1} & 1 &-$\frac{1}{3}$ &-$\frac{1}{3}$
     & 1 &-0.67  &-2.36 \\ [0.5 mm]
     &  &{\bf 8} & 1 &-1 & 0 & 1 &-2  &0 \\ [0.5 mm]  
    & ({\bf 405}, {\bf 1}) &{\bf 1} & 1 &$\frac{5}{33}$ &$\frac{5}{33}$
     & 1 &0.30  &1.07 \\ [0.5 mm]
     &  &{\bf 8} & 1 &$\frac{5}{19}$ & 0 & 1 &0.53  &0  \\ [0.5 mm]
     &  &{\bf 27} & 1 &$\frac{5}{9}$ & 0 & 1 &1.12  &0 
     \\  \hline 
  \end{tabular}
  \end{center}
   \caption{\label{tab8}Ratios of the gaugino masses at the GUT scale
    in the normalization ${M_1}(GUT)$ = 1, and at the electroweak
    scale in the normalization ${M_1}(EW)$ = 1 at the 1-loop level
    for $F$-terms in representations of 
    $SU(3)_C\times SU(3)_L \times U(1) \times SU(2)_X \subset
    SU(6) \times SU(2)_X \subset E_6$.} 
   \end{table}
%

\begin{table}[htb]
\begin{center}
   \begin{tabular}{||c|c|c|ccc|ccc||}
   \hline 
   $E_6$ & $ SU(6)' \times SU(2)_R $ &$SU(3)_L$ & 
                                          $M_1^G$ & $M_2^G$ & $M_3^G$ & 
    $M_1^{EW}$ & $M_2^{EW}$ & $M_3^{EW}$
    \\ \hline 
    {\bf 1} & ({\bf 1}, {\bf 1}) &{\bf 1} &1 &1 &1 &1 &2 &7.1\\ [0.5 mm]
     {\bf 650} & ({\bf 1}, {\bf 1})&{\bf 1} & 1 &-$\frac{5}{13}$ &-$\frac{5}{13}$ 
     & 1 &-0.77 &2.73 \\ [0.5 mm]
     & ({\bf 35}, {\bf 1}) &{\bf 1} & 1 &5 & -5 & 1 &10.0  &-35.5 \\ [0.5 mm]
     &  &{\bf 8} & 1 &-$\frac{5}{3}$ & 0 & 1 &3.35  &0 \\ [0.5 mm]
     & ({\bf 35}, {\bf 3}) &{\bf 1} & 1 &0 & 0 & 1 &0  &0 \\ [0.5 mm]
     &  &{\bf 8} & 1 &0 & 0 & 1 &0  &0 \\ [0.5 mm]
     & ({\bf 189}, {\bf 1}) &{\bf 1} & 1 &-$\frac{5}{3}$ &-$\frac{5}{3}$
     & 1 &-3.35  &-11.8 \\ [0.5 mm]
     &  &{\bf 8} & 1 &5 & 0 & 1 &10.0  &0 \\ [0.5 mm]
  {\bf 2430} & ({\bf 1}, {\bf 1})&{\bf 1} & 1 &$\frac{15}{41}$ &$\frac{15}{41}$  
     & 1 &0.73 &2.59 \\ [0.5 mm]   
     & ({\bf 1}, {\bf 5})&{\bf 1} & 1 &0 &0 & 1 &0 &0 \\ [0.5 mm]   
     & ({\bf 35}, {\bf 3})&{\bf 1} & 1 &0 &0 & 1 &0 &0 \\ [0.5 mm] 
     &  &{\bf 8} & 1 &0 &0 & 1 &0 &0 \\ [0.5 mm]
     & ({\bf 189}, {\bf 1}) &{\bf 1} & 1 &-$\frac{5}{3}$ &-$\frac{5}{3}$
     & 1 &-3.35  &-11.8 \\ [0.5 mm]
     &  &{\bf 8} & 1 &5 & 0 & 1 &10.0  &0 \\ [0.5 mm]
    & ({\bf 405}, {\bf 1}) &{\bf 1} & 1 &$\frac{5}{9}$ &$\frac{5}{9}$
     & 1 &1.12  &3.94 \\ [0.5 mm]
     &  &{\bf 8} & 1 &-$\frac{5}{11}$ & 0 & 1 &-0.91  &0 \\[0.5 mm]
     &  &{\bf 27} & 1 &$\frac{5}{9}$ & 0 & 1 &1.12  &0 \\  \hline 
  \end{tabular}
  \end{center}
   \caption{\label{tab9}Ratios of the gaugino masses at the GUT scale
    in the normalization ${M_1}(GUT)$ = 1, and at the electroweak
    scale in the normalization ${M_1}(EW)$ = 1 at the 1-loop level
    for $F$-terms in representations of 
   $SU(3)_C\times SU(3)_L \times U(1) \times SU(2)_R \subset
    SU(6)' \times SU(2)_R \subset E_6$.}   
  \end{table}
%

\begin{table}[htb]
\begin{center}
\begin{tabular}{||c|c|c|ccc|ccc||}
\hline 
$E_6$ & $ SU(6)'' \times SU(2)_L $ &$SU(3)_R$ & 
                                          $M_1^G$ & $M_2^G$ & $M_3^G$ & 
$M_1^{EW}$ & $M_2^{EW}$ & $M_3^{EW}$
\\ \hline 
{\bf 1} & ({\bf 1}, {\bf 1}) &{\bf 1} &1 &1 &1 &1 &2 &7.1\\ [0.5 mm]
{\bf 650} & ({\bf 1}, {\bf 1})&{\bf 1} & 1 &-5 &1  
& 1 &-10.0 &7.1 \\ [0.5 mm]
& ({\bf 35}, {\bf 1}) &{\bf 1} & 1 &0 & -$\frac{5}{4}$ & 1 &0 &-8.86 \\ [0.5 mm]
& ({\bf 189}, {\bf 1}) &{\bf 1} & 0 &0 &1
& 0 &0  &1 \\ [0.5 mm]
&  &{\bf 8} & 1 &0 & 0 & 1 &0  &0 \\ [0.5 mm]
{\bf 2430} & ({\bf 1}, {\bf 1})&{\bf 1} & 1 &$\frac{35}{9}$ &1  
& 1 &7.81 &7.1 \\ [0.5 mm]    
& ({\bf 189}, {\bf 1}) &{\bf 1} & 0 &0 &1
 & 0 &0  &1 \\ [0.5 mm]
&  &{\bf 8} & 1 &0 & 0 & 1 &0  &0 \\ [0.5 mm]  
& ({\bf 405}, {\bf 1}) &{\bf 1} & 1 &0 &$\frac{5}{12}$
& 1 &0  &2.95 \\ [0.5 mm]
&  &{\bf 8} & 1 &0 & 0 & 1 &0  &0  \\ [0.5 mm]
&  &{\bf 27} & 1 &0 & 0 & 1 &0  &0 
\\  \hline 
\end{tabular}
 \end{center}
\caption{\label{tab10}Ratios of the gaugino masses at the GUT scale
in the normalization ${M_1}(GUT)$ = 1, and at the electroweak
scale in the normalization ${M_1}(EW)$ = 1 at the 1-loop level
for $F$-terms in representations of 
$SU(3)_C\times SU(3)_R \times U(1) \times SU(2)_L \subset
SU(6)'' \times SU(2)_L \subset E_6$.} 
\end{table}

\newpage
 

\end{document}